\documentclass[11pt]{article}
\usepackage[margin=1in]{geometry}
\usepackage[T1]{fontenc}
\usepackage{lmodern}
\usepackage{amsmath,amssymb,amsthm}
\usepackage{enumitem}
\usepackage{booktabs}
\usepackage{array}
\usepackage{graphicx}
\usepackage{tikz}
\usepackage[square,numbers]{natbib}
\usepackage{xcolor}
\usepackage[colorlinks=true,linkcolor=blue!50!black,citecolor=blue!50!black,urlcolor=blue!50!black,backref=page]{hyperref}
\renewcommand*{\backref}[1]{}
\renewcommand*{\backrefalt}[4]{\ifcase #1 \or Cited on page~#2.\else Cited on pages~#2.\fi}
\makeatletter
\def\NAT@alphalabel{}
\newcommand{\natalphalabel}[1]{\gdef\NAT@alphalabel{#1}}
\let\NAT@wrout@num\NAT@wrout
\def\NAT@wrout#1#2#3#4#5{\ifx\NAT@alphalabel\@empty\NAT@wrout@num{#1}{#2}{#3}{#4}{#5}\else\NAT@wrout@num{\NAT@alphalabel}{#2}{#3}{#4}{#5}\fi}
\makeatother

\newtheorem{theorem}{Theorem}
\newtheorem{lemma}[theorem]{Lemma}
\newtheorem{corollary}[theorem]{Corollary}
\newtheorem{proposition}[theorem]{Proposition}

\newtheorem*{informal}{Theorem}
\newcommand{\infthm}{\hyperref[inf:main]{Theorem~(Informal)}}
\theoremstyle{definition}
\newtheorem{definition}[theorem]{Definition}
\newtheorem{example}{Example}
\newtheorem{remark}{Remark}
\newtheorem*{restateddefthm}{\restatedname}
\theoremstyle{plain}
\newcommand{\restatedname}{}
\newtheorem*{restatedthm}{\restatedname}
\newenvironment{restated}[2]{\renewcommand{\restatedname}{#1~\ref{#2}}\begin{restatedthm}[Restated]}{\end{restatedthm}}
\newenvironment{restateddefinition}[1]{\renewcommand{\restatedname}{Definition~\ref{#1}}\begin{restateddefthm}[Restated]}{\end{restateddefthm}}
\newenvironment{restatedproposition}[1]{\begin{restated}{Proposition}{#1}}{\end{restated}}

\newcommand{\alg}{\ensuremath{\textsf{PG}(u)}}
\newcommand{\algmi}{\ensuremath{\textsf{PG}^{-i}}}
\newcommand{\opt}{\textsf{OPT}}
\newcommand{\PG}{\textsf{PG}}
\newcommand{\one}{\mathbf{1}}
\newcommand{\E}{\mathbb{E}}

\newcommand{\HeadU}{0.5915}
\newcommand{\HeadUT}{0.591}
\newcommand{\HeadW}{0.5314}
\newcommand{\HeadWT}{0.531}

\newcommand{\lo}{\underline{u}}
\newcommand{\bz}{\mathbf{z}}
\newcommand{\bu}{\mathbf{u}}

\newcommand{\hi}{\overline{u}}

\title{Online Matching with Arbitrary Stochastic Rewards}
\author{Rajan Udwani\\
UC Berkeley\\
\href{mailto:rudwani@berkeley.edu}{\texttt{rudwani@berkeley.edu}}}
\date{}

\begin{document}
\maketitle
\begin{abstract}
Online matching with stochastic rewards is a generalization of online bipartite matching in which an attempt to match an online arrival to an offline resource may fail with some probability. Each arrival allows at most one attempt and the success probabilities of its edges are revealed when it arrives. Greedy is $0.5$-competitive, which is the best possible for deterministic algorithms. While better competitive ratios are known in several special cases, beating $0.5$ for arbitrary probabilities has remained open. We show that, against the adaptive offline benchmark and for arbitrary probabilities, a randomized Perturbed Greedy algorithm is at least $\HeadUT$-competitive in the unweighted setting and at least $\HeadWT$-competitive with vertex weights. The perturbation functions combine Greedy with scaled versions of the classic exponential function.
\end{abstract}

{\small\setcounter{tocdepth}{1}\tableofcontents}
\section{Introduction}\label{sec:intro}
In the online bipartite matching (OBM) problem introduced by \citet{KVV90}, a set $U$ of offline resources is known in advance and arrivals $t=1,2,\dots,T$ are revealed one at a time. Each arrival must be matched irrevocably upon arrival to at most one available neighbor. Motivated by pay-per-click advertising, \citet{MP12} introduced the problem of online matching with stochastic rewards (OMSR), a generalization of OBM in which an attempted match can fail. Edge $(i,t)$ has a success probability $p_{it}$ revealed when $t$ arrives. Arrival $t$ can be matched by \emph{probing} at most one edge $(i,t)$ to an available resource $i$, and the probe succeeds independently with probability $p_{it}$. Success means $i$ and $t$ are \emph{matched}, which earns a unit reward and makes $i$ unavailable to future arrivals. In vertex-weighted OMSR, a (successful) match of $i$ earns reward $r_i>0$. The objective is to maximize the expected total reward. 

In essence, stochastic rewards are a way of capturing post-allocation uncertainty in online matching. Beyond advertising, such uncertainty manifests in online volunteer matching \citep{MRSS25a,MRSS25b}, gig platforms \citep{Xu24}, online recommendations \citep{GNR14}, and two-sided labor platforms \citep{AS23}.

There are a variety of algorithmic techniques for online matching, and the right choice of technique depends on what information the decision maker has about the arrival sequence. We are interested in the adversarial arrival model, where the decision maker assumes nothing about the arrivals and we compare online algorithms through their competitive ratios. An online algorithm is $\alpha$\emph{-competitive} if its expected reward is at least $\alpha$ times the expected reward of the offline benchmark on every instance. Its \emph{competitive ratio} is the largest such $\alpha$. 

Our overarching goal is to find the algorithm with the optimal competitive ratio for OMSR. For classic OBM, \cite{KVV90} showed that the randomized Ranking algorithm, which fixes a uniformly random order of the resources upfront and matches each arrival to its first available neighbor in this order, is $(1-1/e)$-competitive, and that this is optimal for OBM.  \citet{AGKM11} proposed the Perturbed Greedy algorithm for vertex-weighted OBM and showed that it achieves the same (optimal) guarantee of $1-1/e$. An insightful simplification by \citet{DJK13} showed how these results fall out of a unified primal-dual analysis.

While OMSR is a natural generalization of OBM, which is the special case in which every probe succeeds with probability $1$, the optimal competitive ratio for OMSR remains unknown. Before we dive into this further, it is worth discussing the offline benchmark for OMSR. Ideally, one would hope to compare against a benchmark that knows upfront which edges exist (the outcomes of all probes), but this benchmark is too strong: no online algorithm has a non-trivial guarantee against it. A natural compromise is the benchmark that knows all success probabilities (but not the outcomes) and can probe each arrival at most once, in any order. The outcome of a probe is observed after making it and the probing order may depend on the observed outcomes. This is our benchmark, and we refer to it as the \emph{adaptive offline benchmark}, or simply \emph{offline}. Other benchmarks for this setting are discussed in Section~\ref{sec:related}. Against offline, the best previously known algorithm is Greedy, which probes the available resource with the highest expected reward $r_ip_{it}$. Greedy is $0.5$-competitive for OMSR, even with vertex weights \citep{GNR14,MWZ15}, and it is well known that no deterministic algorithm can do better, even for OBM.\footnote{Consider two resources and two arrivals, with unit weights and probabilities. The first arrival has edges to both resources, and the second arrival has an edge only to the resource that the algorithm matches to the first arrival. A deterministic algorithm matches one arrival, while offline matches both.}

Motivated by applications in online advertising, a stream of work has progressively improved the best known competitive ratio when the probabilities are \emph{vanishing} (asymptotically approaching $0$) \citep{MP12,MWZ15,HZ20,GU23,HJSSWZ23,Udw25}. A few special cases of OMSR, including identical and, more generally, decomposable probabilities ($p_{it}=p_ip_t$ for every edge $(i,t)$), have also been fully resolved \citep{GU23,Udw24}. The problem has also received attention from the other direction, through upper bounds on the competitive ratio \citep{MP12,Udw24,AS25}, including bounds based on hard instances found with reinforcement learning and diffusion models \citep{ZSZJD24,ZSLZYWLD25}. Various generalizations have also been considered, including arrivals with patience \citep{BGMS25,BM25}, online assortment optimization (vertex-weighted, edge-weighted, two-sided) \citep{GNR14,MS20,AS23}, stochastic depletion problems \citep{CF09}, and online submodular welfare maximization with stochastic outcomes \citep{Udw26}. Despite this attention, nothing better than Greedy was known in general, even without vertex weights.

\paragraph{The Main Obstacles.} Beating Greedy requires randomization, and there are two common types of randomized algorithms for (adversarial) online matching: rounding-based algorithms and randomized greedy algorithms. Algorithms of the first type run a fractional online algorithm and round its decisions online. One perspective on this is as follows: in the fractional setting, where the capacity of a resource is consumed in expectation, simple deterministic $(1-1/e)$-competitive algorithms exist \citep{KP00,MSVV07,GNR14}, so one could outperform Greedy by finding a good way to round the fractional solution at each arrival. In a landmark result, \citet{FHTZ22} proposed an online rounding technique called online correlated selection (OCS), which has been applied to many online resource allocation problems, including Adwords \citep{HZZ20}, online matching with stochastic arrivals \citep{TWW22,HSY22}, matching with class fairness \citep{HHIS23}, matching with reusable resources \citep{DFNSU24}, to name a few. In these algorithms the fractional algorithm does not know the realization of the rounding and, in effect, only maintains the probability that each resource has been matched. For OMSR, this separation means that the online rounding itself has to \emph{adapt} to the outcome of each probe. Otherwise the entire algorithm is oblivious to the outcomes, and such \emph{non-adaptive} algorithms, which may probe a resource that has already been matched and earn nothing from the probe, cannot beat $0.5$ even for unweighted OMSR with identical (and vanishing) probabilities \citep[Theorem~5]{MP12}.\footnote{\cite{MP12} show this against the expectation LP benchmark. We show that the same holds against the adaptive offline benchmark (Proposition~\ref{prop:nonadaptive} in Appendix~\ref{app:nonadaptive}).} This prevents a direct (black-box) application of existing OCS and lossless online rounding \citep{BNW23} schemes.\footnote{In existing rounding subroutines, the set of resources from which the subroutine samples may depend on its own past selections. Requiring adaptivity forces this set to depend on the outcomes of the selected probes, and not just on the algorithm's selections. This appears to limit any kind of black-box use of these subroutines, and generalizing them seems (to us) non-trivial.}

Randomized greedy algorithms match greedily according to a randomly perturbed version of the expected reward. Perturbed Greedy (\PG) \citep{AGKM11, GU23} draws a seed $u_i$ uniformly from $[0,1]$, independently for each resource $i$, and probes arrival $t$ with the available resource that maximizes the \emph{reduced price} $r_ip_{it}(1-g(u_i))$, where $g$ is a nondecreasing \emph{perturbation function}:
\begin{flalign*}
&\textbf{Perturbed Greedy:}&&\text{probe arrival $t$ with}\quad\arg\max_{i\text{ available}}\ r_ip_{it}\,\big(1-g(u_i)\big).&&
\end{flalign*} The classical choice is $g(x)=e^{x-1}$. In the unweighted setting with $p_{it}=1$, \PG\ with an increasing $g$ is the Ranking algorithm. \PG\ with the exponential function achieves the optimal guarantee of $1-1/e$ for vertex-weighted OBM \citep{AGKM11} and for OMSR with decomposable probabilities \citep{GU23,Udw24}, and it beats Greedy when the probabilities are vanishing \citep{Udw25}. Therefore, it is a natural candidate for the general setting. But analyzing it in general is challenging and the bottleneck, as identified in prior work \citep{GU23}, is that there is no fixed ranking of the resources. Resource $i$ may be preferred to $j$ at one arrival, and $j$ to $i$ at another.\footnote{A similar crossing property also appears in related \emph{deterministic} problems such as Adwords, where \PG\ is somewhat easier to analyze (see Section~\ref{sec:related} and Remark~\ref{rem:boost}).} 
Since \PG\ is the algorithm we analyze, we discuss this obstacle in more detail in Section~\ref{sec:overview}, along with our main ideas for overcoming it.

\subsection{Our Results}\label{sec:results}
We show that Perturbed Greedy beats $0.5$ for OMSR and arbitrary edge probabilities, against the adaptive offline benchmark. The perturbation functions we use mix Greedy and Ranking and are discussed further in Section~\ref{sec:overview}. Our main results can be summarized as follows.
\phantomsection\label{inf:main}
\begin{informal}[Informal]\leavevmode
\begin{enumerate}[label=\textup{(\Alph*)},ref=\Alph*,leftmargin=*]
\item\label{inf:unweighted} For unweighted OMSR, Perturbed Greedy with a perturbation function that mixes Greedy and Ranking is at least $\HeadUT$-competitive against the adaptive offline benchmark.
\item\label{inf:weighted} For vertex-weighted OMSR, Perturbed Greedy with a (different) perturbation function that mixes Greedy and Ranking is at least $\HeadWT$-competitive against the adaptive offline benchmark.
\end{enumerate}
\end{informal}


Both parts follow from guarantees that we prove for every continuous nondecreasing perturbation function $g$ with $g(1)<1$ (Theorems~\ref{thm:main} and~\ref{thm:weighted} in Section~\ref{sec:alg}). 
When all weights lie within a factor $\Lambda$ of each other, we show a guarantee parameterized by $\Lambda$ that interpolates between the two extremes of $\Lambda=1$ (unweighted) and $\Lambda=\infty$ (arbitrary weights), as shown for a few values of $\Lambda$ in Table~\ref{tab:lambda}. We note that these guarantees are achieved with different perturbation functions (from the mixture family of Section~\ref{sec:model}). For each value of $\Lambda$, the parameters of the function were optimized numerically with AI assistance (see Appendix~\ref{app:numerics}). 

\begin{table}[htbp]
\centering\small
\begin{tabular}{lccccc}
\toprule
weight ratio $\Lambda$ & $1$ (unweighted) & $2$ & $3$ & $10$ & $\infty$ (arbitrary)\\
\midrule
competitive ratio $\ge$ & $0.5915$ & $0.5610$ & $0.5469$ & $0.5358$ & $0.5314$\\
\bottomrule
\end{tabular}
\caption{Lower bounds on the competitive ratio of \PG\ when all weights lie within a factor $\Lambda$ of each other (Corollary~\ref{thm:numbers} in Appendix~\ref{app:numerics}).}\label{tab:lambda}
\end{table}

While our competitive ratio guarantees may not be tight, there is no perturbation function that can achieve a guarantee of $(1-1/e)$ for vertex-weighted OMSR. This result follows by combining prior results of \citet{LTXZZ23} and \citet{Udw24}, and for completeness we include a proof in Appendix~\ref{app:ub}.
\begin{proposition}\label{prop:noopt}
Let $g:[0,1]\to[0,1]$ be nondecreasing. With vertex weights, \PG\ with perturbation function $g$ has competitive ratio at most $(1-1/e-0.0003)$ against the adaptive offline benchmark.
\end{proposition}
In the unweighted setting, no upper bound below $1-1/e$ is known. Overall, this shows that it is possible to beat Greedy for general OMSR, and \PG\ (with the right function) continues to be a powerful algorithm even with arbitrary probabilities and vertex weights. Finding the optimal guarantee and algorithm for OMSR remains open (along with various other follow-up questions that we discuss in Section~\ref{sec:conclusion}). 

\subsection{Technical Overview}\label{sec:overview}
When the probabilities are decomposable, \PG\ reduces to randomly ranking resources according to $r_ip_i(1-g(u_i))$. If a resource $i$ is preferred to $j$ at one arrival, then it is preferred at every arrival. This structure breaks down in general, and to understand how this impacts the analysis, we recall the classic primal-dual analysis of \cite{DJK13} (which \citet{GU23} generalized for OMSR). The main idea is to define a dual fitting procedure and show that in expectation the solution it generates is approximately dual feasible. Formally, consider the matching LP of vertex-weighted OBM and its dual:
\[
\begin{array}[t]{@{}l@{\quad}l@{\ \ }l@{\quad}l@{}}
\text{(Primal)} & \max & \sum_{(i,t)\in E}r_ix_{it} & \\
 & \text{s.t.} & \sum_tx_{it}\le1 & \forall i\in U,\\
 & & \sum_ix_{it}\le1 & \forall t\in[T],\\
 & & x_{it}\ge0 & \forall(i,t)\in E;
\end{array}
\qquad
\begin{array}[t]{@{}l@{\quad}l@{\ \ }l@{\quad}l@{}}
\text{(Dual)} & \min & \sum_{i\in U}\theta_i+\sum_{t\in[T]}\lambda_t & \\
 & \text{s.t.} & \lambda_t+\theta_i\ge r_i & \forall(i,t)\in E,\\
 & & \theta_i\ge0 & \forall i\in U,\\
 & & \lambda_t\ge0 & \forall t\in[T].
\end{array}
\]
A dual fitting constructs random duals whose total equals the value of the algorithm for every realization of its randomness, and which satisfy every dual constraint in expectation up to a factor $\alpha$. Weak duality then shows that the algorithm is $\alpha$-competitive. For OMSR, we follow the LP-free approach of \citet{GIU}. For simplicity, in this section we convey the main ideas through the classic dual LP described above.

Given a random seed vector $\bu=(u_i)$, \cite{DJK13} propose the following dual fitting: when \PG\ matches $i$ at $t$, arrival $t$ gets the price $\lambda_t(\bu)=r_i(1-g(u_i))$ and resource $i$ gets the rest, $\theta_i(\bu)=r_ig(u_i)$. Since $i$ is matched at most once, we can also write $\theta_i(\bu)=\sum_t\theta_{it}(\bu)$, where $\theta_{it}(\bu)=r_ig(u_i)$ if \PG\ matches $i$ at $t$, and $\theta_{it}(\bu)=0$ otherwise. They show that, in expectation, $\E[\lambda_t(\bu)+\theta_i(\bu)]\ge(1-1/e)\,r_i$ for every edge $(i,t)$. Analyzing the high-dimensional expectations $\E[\lambda_t(\bu)]$ and $\E[\theta_i(\bu)]$ is challenging because the matching can change completely with just a few changes to the seeds. \cite{DJK13} fix the seeds $\bu_{-i}$ of all resources other than $i$ and take the expectation with respect to $u_i$ alone, proving a stronger statement that holds for arbitrary fixed seeds of the other resources. 

When $u_i=1$ (and $g(1)=1$, as for the classical $g(x)=e^{x-1}$), the reduced price of $i$ is $0$ everywhere, so $i$ is essentially removed from the algorithm. As $u_i$ decreases, they show that $\lambda_t(\bu)$ never drops below its value at $u_i=1$, because adding $i$ can only enlarge the set of resources available at any arrival. \cite{GU23} showed that an appropriately generalized version of this is true even for OMSR with arbitrary probabilities. Their analysis conditions not only on the seeds of the other resources but also on the outcomes of the other edges, i.e., it works on a fixed \emph{sample path}. 

\begin{figure}[htbp]
\centering
\begin{tikzpicture}[
  arr/.style={draw,rectangle,minimum size=6mm,inner sep=1pt},
  res/.style={draw,circle,minimum size=7mm,inner sep=0pt},
  pg/.style={very thick},
  pgmi/.style={thick,dashed},
  lab/.style={fill=white,inner sep=1pt,font=\footnotesize,text=black},
  faded/.style={draw=black!25,text=black!35}]
\begin{scope}
\node[res,faded] (ia) at (0,2) {$i$};
\node[font=\scriptsize,text=black!45] at (0,2.6) {deleted ($u_i=1$)};
\node[res] (ja) at (2,2) {$j$};
\node[arr] (a1a) at (0,0) {$1$};
\node[arr] (a2a) at (2,0) {$2$};
\draw[pgmi] (ja) -- node[lab,pos=0.5] {$1/2$\ \checkmark} (a1a);
\node[align=center,font=\footnotesize] at (1,-0.95) {without $i$:\\ matches $j$ at $1$,\\ probes nothing at $2$};
\end{scope}
\begin{scope}[shift={(5,0)}]
\node[res] (ib) at (0,2) {$i$};
\node[res] (jb) at (2,2) {$j$};
\node[arr] (a1b) at (0,0) {$1$};
\node[arr] (a2b) at (2,0) {$2$};
\draw[pg] (ib) -- node[lab,pos=0.5] {$1/4$\ $\times$} (a1b);
\draw[pg] (jb) -- node[lab,pos=0.5] {$1$\ \checkmark} (a2b);
\node[align=center,font=\footnotesize] at (1,-0.95) {with $i$:\\ the probe of $i$ at $1$ fails,\\ $j$ is matched at $2$};
\end{scope}
\end{tikzpicture}
\caption{The runs of \PG\ in Example~\ref{ex:release} on a sample path on which $(j,1)$ and $(i,2)$ succeed and $(i,1)$ fails (\checkmark: the probe succeeds; $\times$: it fails). Squares are arrivals and circles are resources; the label of an edge is its probability. Left: the run without $i$. Right: the run with $i$, which releases $j$ at arrival~$1$ and never matches $i$.}\label{fig:release}
\end{figure}
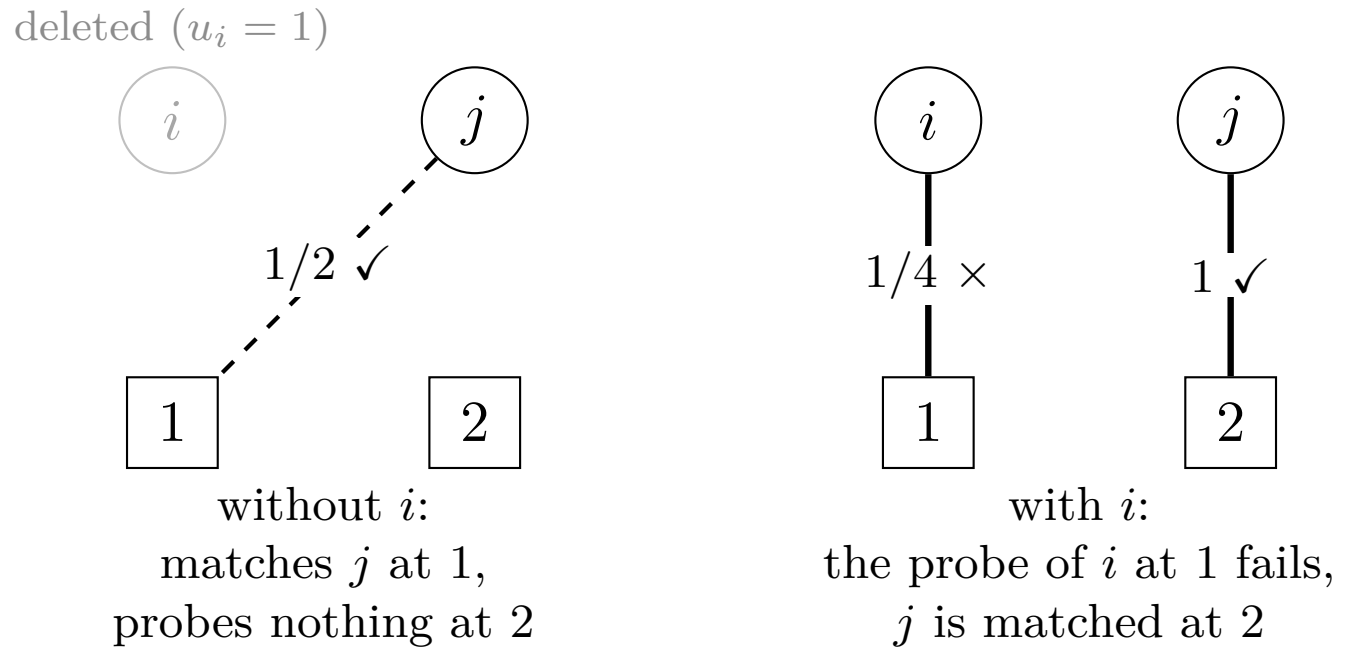

To lower bound $\E[\theta_i(\bu)\mid \bu_{-i}]$, \cite{DJK13} showed that if the resource $i$ is matched at some seed $u_i=u_0$, then it continues to be matched for all smaller values of $u_i$, because $i$ becomes more attractive as its rank improves ($u_i$ decreases). As the following example (similar to an example of \cite{GU23}) demonstrates, this is not true for OMSR: on a fixed sample path, $i$ may be matched for one seed but not for a better seed. 
\begin{example}[A Release]\label{ex:release}
Consider an unweighted instance with two resources $i$ and $j$ and two arrivals. Arrival~$1$ has edges to $i$ and $j$ with $p_{i1}=1/4$ and $p_{j1}=1/2$, and arrival~$2$ has edges to $i$ and $j$ with $p_{i2}=p_2$ and $p_{j2}=1$, where $p_2\in(0,1/2)$. Focus on a sample path on which the edges $(j,1)$ and $(i,2)$ succeed and the edge $(i,1)$ fails. The run of \PG\ on the instance without $i$ matches $j$ at arrival~$1$ and probes nothing at arrival~$2$. When $i$ is poorly ranked so that $g(u_i)>g(u_j)$, \PG\ matches $j$ at arrival~$1$, and then matches $i$ at arrival~$2$. With a better seed $u_i$ such that
\[
p_2\,(1-g(u_i))\ <\ 1-g(u_j)\ <\ \tfrac12\,(1-g(u_i)),
\]
$i$ beats $j$ at arrival~$1$, and $j$ beats $i$ at arrival~$2$. Then \PG\ instead probes $i$ at arrival~$1$; the probe fails and \emph{releases} $j$, which would have been matched at arrival~$1$ without $i$. Because $j$ beats $i$ at arrival~$2$, \PG\ matches $j$ there, and $i$ does not get another chance (Figure~\ref{fig:release}). So $\theta_i(\bu)=0$ on this sample path, while the offline benchmark can match $j$ at arrival~$1$ and $i$ at arrival~$2$.
\end{example}

This example demonstrates a potential limitation of the conditional analysis and not of \PG\ itself. A natural remedy is to remove the conditioning on the seeds of the other resources and on the outcomes of the other edges, so that the sample paths on which $i$ is matched can pay for those on which it is not. However, an analysis without conditioning remains particularly difficult because, in addition to taking the expectation with respect to $\bu$, we also need to take the expectation with respect to the outcomes of the probes. There may also be long chains of releases (Example~\ref{ex:chain} in Section~\ref{sec:weighted}), so taking the expectation with respect to just a few seeds or a few edges may not suffice. We explore a different path that retains much of the tractability of the conditional analysis. First, we reformulate the problem, and then we modify the algorithm and the analysis.
\paragraph{Correlated Edges.} Instead of working with independent outcomes, we use a simple coupling that correlates the outcomes of the edges incident on the same arrival. For each arrival $t$, draw a threshold $z_t$ uniformly from $[0,1]$, and let the edge $(i,t)$ succeed iff $z_t\le p_{it}$. Every algorithm (online or offline) probes at most one edge at each arrival, so this is an equivalent reformulation of OMSR (formalized in Section~\ref{sec:model}). In Example~\ref{ex:release}, this means that if $j$ is (successfully) matched at arrival~$1$ in the run without $i$, then $z_1\le p_{j1}=1/2$, and $i$ is (successfully) matched at arrival~$1$ in the run with $i$ with conditional probability $p_{i1}/p_{j1}=1/2$, instead of $p_{i1}=1/4$ with independent edges. One way to view this is as a coupling between the runs with and without $i$, and it allows us to continue using conditional expectations while reaping some of the potential gains from taking the expectation over the randomness of the probes.

\paragraph{Mix in Greedy.} We make a small but helpful modification to the algorithm by mixing Greedy into Ranking. This is rooted in two observations. First, Greedy is weak when there are ties, i.e., when the online algorithm does not know which of several resources to pick at an arrival, and in this case randomization, and in particular Ranking, helps (this is the classic insight of \cite{KVV90}). But Greedy is good when we are far away from ties.\footnote{If at every arrival the expected rewards of any two edges differ by a factor of at least $1+\eta$, then Greedy is $\frac{1+\eta}{2+\eta}$-competitive, i.e., $(0.5+\Theta(\eta))$-competitive for $\eta\le1$ and asymptotically optimal as $\eta\to\infty$ (Appendix~\ref{app:greedy}). 
} In unweighted scenarios such as Example~\ref{ex:release}, either the probabilities of $(i,1)$ and $(j,1)$ are close, in which case the coupling helps a lot by boosting the (conditional) success probability of $(i,1)$, or else the probabilities are far apart ($p_{i1}\ll p_{j1}$), in which case a greedy selection at arrival~$1$ picks $j$, which is a good call. In general, since the perturbation factor $1-g(u_j)$ of a resource is the same at every arrival, crossings can only arise if there is variation in the probabilities. Second, mixing Greedy into Ranking does not fundamentally alter the algorithm, because there are perturbation functions $g$ that capture the mixture.
To see this, write the perturbation factor of a resource with seed $u$ as
\begin{equation}\label{eq:additive}
	1-g(u)\;=\;\underbrace{\big(1-g(1)\big)}_{\text{greedy part}}\;+\;\underbrace{\big(g(1)-g(u)\big)}_{\text{Ranking part}} .
\end{equation}
When $g(1)<1$, the reduced price $r_ip_{it}(1-g(u_i))$ is then the sum of a \emph{greedy price} $r_ip_{it}(1-g(1))$, which every resource keeps whatever its seed, and a \emph{Ranking perturbation} $r_ip_{it}(g(1)-g(u_i))$. 
More generally, we allow a randomized mixture, in which at the start each resource independently decides whether to follow a greedy price or a Ranking-based reduced price. The perturbation functions of Section~\ref{sec:model} capture both types of mixture.
\paragraph{Modifications to Analysis.} We make a few observations based on Example~\ref{ex:release} to give a high-level idea of the main changes. In the example, $j$ is released (and $i$ is unmatched) whenever $1/4<z_1\le1/2$, even with correlated edges. First, notice that the release of $j$ has a positive effect on the reduced price at arrival~$2$, which is at least $p_{i2}(1-g(u_i))$ \emph{regardless} of whether $i$ is available at the arrival. Thus, while the release depresses $\theta_i$, it offers some compensation in the form of a boost to $\lambda_2$, even on sample paths where $i$ is unavailable at arrival~2. Second, there is a big imbalance between the sample paths on which $i$ is matched at arrival~$1$ and those on which it is not. Therefore, it is reasonable to equalize the value of $\theta_i=\sum_t\theta_{it}$ across these paths. At arrivals $t$ where such a release could cause trouble later, we spread $\theta_{it}$ evenly over $z_t\leq p_{jt}$, so that the sample paths on which $i$ is unsuccessful at $t$ but releases a resource $j$ (which may prevent further attempts on $i$) get some gain transferred from the paths on which $i$ is successful. At the other arrivals, we keep $\theta_{it}$ as it is on each sample path. In the unweighted setting, the greedy part of the reduced price ensures that this equalization passes a sufficiently large value to the paths on which $i$ is unsuccessful at arrival~$1$. Since \PG\ prefers $i$ to $j$ at arrival~$1$,
\[
p_{i1}\,(1-g(u_i))\ \ge\ p_{j1}\,(1-g(u_j))\ \ge\ p_{j1}\,(1-g(1))\qquad\Rightarrow\qquad \frac{p_{i1}}{p_{j1}}\ \ge\ \sigma(u_i):=\frac{1-g(1)}{1-g(u_i)} .
\]
The greedy part $1-g(1)$, which every resource keeps whatever its seed, prevents \PG\ from preferring a resource that is much less likely to succeed. In other words, if $i$ is selected at arrival~$1$, then its success probability cannot be much lower than that of $j$. So spreading $\theta_{i1}$, which is $g(u_i)$ if $z_1\le p_{i1}$ and $0$ otherwise, evenly over the outcomes $z_1\le p_{j1}$ leaves at least $\sigma(u_i)g(u_i)$ on each of them (Figure~\ref{fig:avg} in Section~\ref{sec:relarr}).


These are the key ideas in the unweighted setting. However, with arbitrary vertex weights, it is possible that $p_{i1}\ll p_{j1}$ but $r_i\gg r_j$, and even Greedy prefers	 $i$ to $j$ at arrival~$1$ but switches to $j$ at arrival~$2$. Then the ratio $p_{i1}/p_{j1}$ can be arbitrarily small, and the averaging is not by itself strong enough. Here we observe that the crossing inequalities have more to say. Suppose for simplicity that $p_{j1}=1$, but $i$ still beats $j$ at arrival~$1$, and $j$ beats $i$ at arrival~$2$. Then the reduced prices satisfy
\[
p_{i2}\,r_i(1-g(u_i))\ <\ r_j(1-g(u_j))\ <\ p_{i1}\,r_i(1-g(u_i)),\qquad\text{so}\qquad p_{i2}\ <\ p_{i1} .
\]
So the edge to $i$ at arrival~$2$ succeeds with probability at most the probability $p_{i1}$ with which \PG\ matches $i$ at arrival~$1$. We can therefore split $\theta_i$ into two halves, so that each of the two edges to $i$ receives $r_ig(u_i)/2$ whenever it succeeds; in expectation this costs $(p_{i1}+p_{i2})\,r_ig(u_i)/2\le p_{i1}\,r_ig(u_i)=\E[\theta_i]$. In general, the example extends to a long chain (Example~\ref{ex:chain} and Figure~\ref{fig:chain} in Section~\ref{sec:weighted}): the competitor that the first failed probe of $i$ releases beats $i$ at the second arrival, this releases another resource that beats $i$ at the third arrival, and so on. Along such a chain the success probabilities of $i$ will shrink geometrically because at every arrival $i$ loses to a resource that it beat at the previous arrival. So we split $\theta_i$ geometrically along the chain, giving a proportionally larger share to arrivals where the reduced price of $i$ is significantly higher than that of the resource selected at this arrival in the run without $i$ (signaling a bigger loss compared to the classic analysis if $i$ is not matched at this arrival). 
We show that such a split is always possible in expectation, on every instance. This gives a new lower bound on what the duals pay, which is insufficient on its own, but gives the result for arbitrary weights when combined with the generalization of the unweighted lower bound. 

\subsection{Related Work}\label{sec:related}
There is a vast literature on online resource allocation. We focus on closely related work and refer to the surveys and tutorials of \citet{Meh13}, \citet{HTW24}, \citet{BBP24} and \citet{NU26} for other related work. Unless stated otherwise, the results for OMSR that we mention are for the vertex-weighted setting.

\paragraph{Deterministic Rewards.} We start with OBM, the deterministic special case of OMSR, where the offline benchmark is simply a maximum (weight) matching. \cite{KVV90} showed that randomly ranking the resources gives the optimal guarantee of $1-1/e$,\footnote{\citet{BM08} offered a simplified combinatorial proof.} while breaking ties independently at random at each arrival is only $0.5$-competitive like Greedy. \citet{KP00} showed that for $b$-matching, where each resource can be matched to $b$ arrivals, a deterministic water-filling algorithm is $(1-1/e)$-competitive when the capacity $b$ is large, which is best possible even in this setting. Inspired by search advertising, \citet{MSVV07} introduced the Adwords problem, a budgeted generalization in which each resource has a budget and a bid for each arrival. Matching an arrival to a resource earns the bid, subject to total earnings of at most the budget of each resource. They proposed an algorithm that matches each arrival greedily according to the perturbed bid $b_{it}(1-g(x_i))$, where $x_i$ is the fraction of the budget of $i$ that has been spent, and showed that setting $g(x)=e^{x-1}$ gives the optimal competitive ratio of $1-1/e$ when the budgets are much larger than the individual bids (the small bids assumption, which generalizes large capacities). \cite{AGKM11} considered vertex-weighted OBM (with unit capacities) and introduced \PG, showing that the same exponential function is $(1-1/e)$-competitive. \citet{Vaz23} and \citet{AS25b} strengthened this result, showing that \PG\ with a single seed for each resource is $(1-1/e)$-competitive also for vertex-weighted $b$-matching with arbitrary capacities. \citet{BJN07} introduced the online primal-dual approach in the context of online resource allocation, and \cite{DJK13} gave the first primal-dual analysis of Ranking and \PG. This was further simplified by \citet{EFFS21}.
\paragraph{The Expectation LP Benchmark.} \cite{MP12} introduced OMSR and proposed a deterministic LP relaxation of the offline benchmark, which we earlier referred to as the expectation LP. This is the optimal value of the fractional budgeted allocation problem in which resources have unit budget and a match of $i$ to $t$ earns $r_ip_{it}$ and uses the fraction $p_{it}$ of $i$'s budget. For the special case of unweighted OMSR with identical vanishing probabilities, they showed that a deterministic algorithm is $0.567$-competitive, the randomized Ranking algorithm is $0.534$-competitive, and that no algorithm is better than $0.621$-competitive. Their analyses introduced the idea of viewing each resource as having a random budget that is revealed only when the ``load'' of the resource reaches it, like Adwords with unknown (random) budgets. 
\cite{MP12} also showed that adaptivity is necessary: no non-adaptive algorithm is better than $0.5$-competitive against this benchmark (in fact, their instance implies this bound against the offline benchmark, see Appendix~\ref{app:nonadaptive}). For unweighted OMSR, \citet{MWZ15} gave a $0.534$-competitive algorithm for non-identical vanishing probabilities, and showed that a non-adaptive version of Greedy is $0.5$-competitive in general. \citet{HZ20} proposed a configuration LP that converges to the expectation LP as the probabilities vanish, and used it in a primal-dual analysis to obtain a deterministic $0.576$-competitive algorithm for identical vanishing probabilities and a randomized $0.572$-competitive algorithm for non-identical vanishing probabilities. \citet{HJSSWZ23} improved the analysis of Ranking for unweighted OMSR with identical (not necessarily vanishing) probabilities from $0.534$ to $0.572$. When resources have large capacities, \citet{GNR14} generalized the algorithm of \cite{MSVV07} and showed that it is $(1-1/e)$-competitive for a generalization of OMSR. In fact, even non-adaptive deterministic algorithms achieve this guarantee \citep{GIU,Udw26}. Using machine learning to search for hard instances, \citet{ZSZJD24} and \citet{ZSLZYWLD25} improved the upper bound of \cite{MP12} to $0.597$ and $0.594$, respectively, for identical vanishing probabilities.

\paragraph{The Adaptive Offline Benchmark (and LP relaxations).} Recall that the adaptive offline benchmark knows all success probabilities and may probe the arrivals in any order, adapting to the observed outcomes. The expectation LP is an upper bound on this benchmark. \citet{GU20,GU23} proposed a natural generalization of \PG\ for vertex-weighted OMSR and showed that it is $(1-1/e)$-competitive when the probabilities are decomposable. Since this includes identical probabilities, the upper bound of $0.621<1-1/e$ of \cite{MP12} is an artifact of the expectation LP being a loose upper bound. They also showed that a deterministic algorithm is $(1-O(\sqrt{p}))\,0.596$-competitive, where $p$ is the largest edge probability (using the random budget view of \cite{MP12}). This beats Greedy for sufficiently small constant probabilities and approaches $0.596$ as the probabilities vanish. \cite{Udw24} simplified the analysis for decomposable probabilities through a reduction to deterministic rewards. This reduction also couples the outcomes of edges but with a different coupling from the one we use: when $p_{it}=p_ip_t$, a common random variable decides whether arrival $t$ exists (with probability $p_t$), and the remaining success probabilities $p_i$ depend only on the resources. \cite{Udw24} also showed that \PG\ with the exponential function is at most $0.624$-competitive for vertex-weighted OMSR.

\paragraph{The Clairvoyant Benchmark (and LP relaxations).} This is a fixed order offline benchmark, weaker than the adaptive offline benchmark because of the restriction that it must select edges to probe in the arrival order (same as an online algorithm). Prior to \cite{MP12}, \citet{CF09} introduced a general framework of stochastic depletion problems that includes OMSR as a special case, and showed that Greedy is $0.5$-competitive against the clairvoyant benchmark in this generality. \cite{HJSSWZ23} proposed a configuration LP relaxation of the clairvoyant benchmark, and showed that, in the unweighted setting, a deterministic algorithm is $0.611$-competitive for vanishing probabilities and $0.613$-competitive for identical vanishing probabilities. 

We note that the upper bound of $1-1/e$ for OBM \citep{KVV90} does not directly apply to vanishing probabilities against the offline and clairvoyant benchmarks. \citet{AS25} bridged this gap and proved that no algorithm is better than $(1-1/e)$-competitive against the clairvoyant (and therefore the offline) benchmark even for identical vanishing probabilities.

\paragraph{Adwords with Unknown Budgets.} With vanishing probabilities, OMSR is a special case of Adwords with small bids, in which the budget of each resource is drawn from an exponential distribution with mean $1$ and is revealed to the algorithm only when it is used up \citep{MP12,HZ20,Udw25}. \citet{Udw25} studied Adwords with adversarial unknown budgets and showed that a budget-oblivious version of \PG, which matches each arrival to the resource with the largest perturbed bid $b_{it}(1-g(u_i))$, is $0.522$-competitive for a scaled exponential function. (This implies the same guarantee for OMSR with vanishing probabilities.) 
Crossings (resource $i$ preferred to $j$ at one arrival and $j$ to $i$ at another) also arise in this setting, because the bids of a resource vary across arrivals. However, the absence of stochastic outcomes allows for a simple replacement for the missing lower bound on $\theta_i$: the troublesome case is simply when $i$ is available at every arrival, and a comparison between the total reduced prices in the runs with and without $i$ gives a lower bound on $\theta_i$ in this case. \cite{LTXZZ23} study how the perturbation function affects \PG\ for vertex-weighted OBM and Adwords. They show that \PG\ is not $(1-1/e)$-competitive for fractional Adwords for any perturbation function. In Appendix~\ref{app:ub}, we transfer this upper bound to vertex-weighted OMSR.

\paragraph{Outline.} Section~\ref{sec:model} formally defines the model, the correlated model, and Perturbed Greedy. Section~\ref{sec:alg} states our results formally, defines the coupling and the candidate dual solution, reduces the competitive ratio to a per-seed inequality. Section~\ref{sec:analysis} proves the unweighted per-seed inequality and explains why the argument does not extend directly to weights. Section~\ref{sec:weighted} proves the weighted per-seed inequality. Section~\ref{sec:conclusion} discusses open problems, including models beyond OMSR. Appendix~\ref{app:numerics} gives the perturbation functions we use and their guarantees, Appendix~\ref{app:ub} the upper bound on \PG, Appendix~\ref{app:nonadaptive} the bound for non-adaptive algorithms, and Appendix~\ref{app:greedy} the guarantee of Greedy away from ties. Appendix~\ref{app:deferred} contains deferred proofs, and Appendix~\ref{app:Ip} the unweighted argument with weights, including the proof of Lemma~\ref{prop:Ip}(A).

\section{Correlated Model and Preliminaries}\label{sec:model}

\paragraph{Setting.} An instance of vertex-weighted OMSR consists of a finite set $U$ of resources, arrivals $t\in[T]=\{1,\dots,T\}$ arriving in this order, a set of edges $E\subseteq U\times[T]$ with success probabilities $p_{it}\in(0,1]$ for $(i,t)\in E$, and a weight $r_i>0$ for each resource $i\in U$. Arrival $t$ and the probabilities $\{p_{it}\}_{i\in U}$ are revealed at time $t$\footnote{Alternatively, consider a complete bipartite graph in which some edges may have probability $0$. We exclude such edges from $E$ so that $1/p_{it}$ is always defined.}. Each resource has unit capacity\footnote{The setting with arbitrary capacities reduces to unit capacity by duplicating resources.}. The unweighted setting is the special case where $r_i=1$ for every resource $i\in U$. 

\paragraph{(Non-Anticipative) Online and Offline Policies.} A (possibly randomized) non-anticipative policy maintains the set of available resources and the set of arrivals probed so far. At each step it either stops or \emph{probes} an edge $(i,t)\in E$, that is, attempts to match $t$ to $i$, where $i$ is available and $t$ has not been probed. The choice may depend on the internal randomness of the policy and on the outcomes of its previous probes \emph{but not on the success of} $(i,t)$. The policy then observes the outcome. A successful probe is a \emph{match}: it yields value $r_i$ and consumes $i$ (resources have unit capacity). A failed probe leaves the resource unused. The arrival $t$ is marked as probed in either case. 

An \emph{online} policy must in addition probe arrivals in increasing order and decide on $t$ before seeing later arrivals. The benchmark $\opt$ is an optimal non-anticipative policy, with no ordering restriction. Without loss of generality, $\opt$ is deterministic and never probes an unavailable resource\footnote{An optimal non-anticipative policy can be computed by a (high-dimensional) dynamic program. Selecting an unavailable resource or randomizing over actions cannot increase the expected reward.}. Overloading notation, $\E[\opt]$ denotes the expected total reward of $\opt$.

\paragraph{Independent and Correlated Models.} In the standard (independent) model each edge $(i,t)$ succeeds independently with probability $p_{it}$. In the \emph{correlated} model we draw thresholds $\bz=(z_t)_{t\in[T]}$ i.i.d.\ uniform on $[0,1]$, independent of everything else, and declare edge $(i,t)$ successful iff $z_t\le p_{it}$. So if edge $(i,t)$ succeeds, then every edge $(j,t)$ with $p_{jt}\ge p_{it}$ also succeeds. The two models are equivalent for every non-anticipative policy. We state this formally and include the proof in Appendix~\ref{app:deferred}.

\begin{lemma}\label{lem:transform}
For every non-anticipative policy, the joint law of its sequence of probes and outcomes, and hence the law of its value, is the same in the independent and the correlated models.
\end{lemma}

From now on we work in the correlated model. We write 
\[\bz_{-t}=(z_s)_{s\ne t}\quad \text{ and }\quad \bz_{<t}=(z_s)_{s<t}.\] 
A deterministic non-anticipative policy decides whether to probe $(i,t)$ before it observes the outcome of any edge of $t$, so whether it probes $(i,t)$ is determined by $\bz_{-t}$ and does not depend on the threshold $z_t$.

\paragraph{Perturbed Greedy (\PG).} Fix a continuous nondecreasing function $g:[0,1]\to[0,1]$ with $g(1)<1$. \PG\ draws a seed $u_i\sim U[0,1]$ independently for each resource $i$, and we let $\bu=(u_i)_{i\in U}$ denote the seed vector. Upon arrival of $t$, \PG\ considers the set $A_t$ of available resources with an edge to $t$ and probes the resource with the maximum \emph{reduced price}
\[
P_t=\max_{i\in A_t}\ r_ip_{it}(1-g(u_i)),
\]
breaking ties in a fixed order of the resources. 
We set $P_t=0$ when $A_t=\emptyset$. 

\paragraph{The Mixture Family.} 
As discussed in Section~\ref{sec:overview}, we mix Greedy into Ranking in two ways. In a \emph{deterministic} mixture, the perturbation function is a convex combination of a greedy part and a Ranking part (as previously discussed). In a \emph{randomized} mixture, each resource is independently priced greedily (without any perturbation) with some probability, and priced according to the exponential (Ranking) perturbation function otherwise. We consider a family of functions that captures both types of mixtures. To compare unperturbed and perturbed resources, we apply a fixed (deterministic) multiplier to the expected reward of unperturbed resources and a randomized perturbation to that of perturbed resources. A natural way to do so is to set aside a band of seeds for the scenario where the resource is unperturbed.
Fix a band $[\lo,\hi]\subseteq[0,1]$ of seeds that represent Greedy. This band is effectively removed from the seed, so that the effective seed of a perturbed resource is
\[
s(x)=\min\{x,\lo\}+\max\{0,x-\hi\}.
\]
Note that this function is constant on the band. The \emph{mixture family} consists of the functions
\begin{equation}\label{eq:sband}
g(x)=A\,e^{k\,s(x)}+B,
\end{equation}
with $A,k>0$ and $B\ge0$ such that $g(1)<1$. A resource whose seed falls in the band, which happens with probability $\hi-\lo$, has the constant multiplier $1-g(\lo)$, and among such resources \PG\ compares $r_ip_{it}$ alone, i.e., it prices them greedily. A resource whose seed lies outside the band is priced by a scaled Ranking function of its effective seed. The band captures randomized mixtures, and the offset $1-g(1)>0$ captures the deterministic one. The full band $[0,1]$ is Greedy, and the empty band with $A=e^{-1}$, $k=1$ and $B=0$ is the classical exponential function.


\paragraph{Deterministic Tie Breaking.}  Ties can occur with non-zero probability when $g$ has flat pieces. Breaking them in a fixed order of the resources implies the following property, 
\begin{quote}
	\emph{If \PG\ selects the resource $k$ from an available set $A$, and $k\in B\subseteq A$, then \PG\ also selects $k$ from the available set $B$.}
\end{quote}
We use this property to compare different runs of \PG\ in several lemmas. 

\section{Statements of Main Results and the Dual Certificate}\label{sec:alg}

Throughout, the perturbation function $g:[0,1]\to[0,1]$ is continuous and nondecreasing with $g(1)<1$, as in Section~\ref{sec:model}. Let
\[
\sigma(x):=\frac{1-g(1)}{1-g(x)}.
\]
Our competitive ratio guarantee in the unweighted setting is
\begin{equation}\label{eq:alpha}
\alpha(g):=\min\Big\{\alpha_1(g),\ \min_{y\in[0,1]}\alpha_2(g,y)\Big\},
\end{equation}
where
\begin{align*}
\alpha_1(g)&:=\int_0^1\min\{g(x),\ \sigma(x)g(x)+1-g(x)\}\,dx,\\
\alpha_2(g,y)&:=1-g(y)+\int_0^y\min\{g(x),\ \sigma(x)g(x)+g(y)-g(x)\}\,dx.
\end{align*}
\begin{theorem}\label{thm:main}
	Perturbed Greedy with perturbation function $g$ is at least $\alpha(g)$-competitive against the adaptive offline benchmark for unweighted OMSR.
\end{theorem}
The first term $\alpha_1(g)$ would be subsumed by the second term if we had $g(1)=1$, since then $\alpha_1(g)=\alpha_2(g,1)$. Comparing the second term with the classical one for OBM \citep{DJK13}, where the competitive ratio can be expressed as $1-g(y)+\int_0^yg(x)\,dx$, the function $g(x)$ in the integrand is replaced by
\[
\min\{g(x),\ \sigma(x)g(x)+g(y)-g(x)\}\ \le\ g(x).
\]



In the vertex-weighted setting, let $\Lambda\in[1,\infty]$ be an upper bound on the ratio $r_i/r_j$ of the weights of any two resources $i,j$, where $\Lambda=\infty$ allows arbitrary weights. Let $\sigma_\Lambda(x):=\sigma(x)/\Lambda=(1-g(1))/(\Lambda(1-g(x)))$. Let $c:[0,1]\to[0,1]$ be an integrable function with $0\le c\le g$, which we call the \emph{tradeoff function}. Define
\begin{equation}\label{eq:alphawL}
\alpha_\Lambda(g,c):=\min\Big\{\alpha_{\Lambda,1}(g,c),\ \min_{y\in[0,1]}\alpha_{\Lambda,2}(g,c,y)\Big\},
\end{equation}
where
\begin{align*}
\alpha_{\Lambda,1}(g,c)&:=\int_0^1\big(\min\{1-g(x)+(g(x)-c(x))\sigma_\Lambda(x),\ g(x)-c(x)\}+c(x)\big)\,dx,\\
\alpha_{\Lambda,2}(g,c,y)&:=1-g(y)+\int_0^y\Big(\min\{g(y)-g(x)+(g(x)-c(x))\sigma_\Lambda(x),\ g(x)-c(x)\}\\
&\qquad\qquad+c(x)\,\min\Big\{1,\ \frac{g(y)-g(x)}{(1-g(x))(1-\sigma_\Lambda(x))}\Big\}\Big)\,dx\qquad(\text{read as $1$ if $\sigma_\Lambda(x)=1$}).
\end{align*}
For arbitrary weights, $\sigma_\infty\equiv0$ and the guarantee is
\begin{multline}\label{eq:alphaw}
\alpha_\infty(g,c)=\min\Big\{\int_0^1\big(\min\{ g(x)-c(x),\ 1-g(x)\}+c(x)\big)\,dx,\\
\min_{y\in[0,1]}\Big[1-g(y)+\int_0^y\Big(\min\big\{ g(x)-c(x),\ g(y)-g(x)\big\}+c(x)\,\frac{g(y)-g(x)}{1-g(x)}\Big)\,dx\Big]\Big\}.
\end{multline}
\begin{theorem}\label{thm:weighted}
Let $\Lambda\in[1,\infty]$. For every $c$ with $0\le c\le g$, Perturbed Greedy with perturbation function $g$ is at least $\alpha_\Lambda(g,c)$-competitive against the adaptive offline benchmark for vertex-weighted OMSR on instances whose weights lie within a factor $\Lambda$ of each other. In particular, it is at least $\alpha_\infty(g,c)$-competitive for arbitrary weights.
\end{theorem}
The guarantee interpolates between the unweighted and (arbitrary) weighted setting. For $\Lambda=1$ and $c\equiv0$ it is $\alpha(g)$, so Theorem~\ref{thm:weighted} generalizes Theorem~\ref{thm:main}.  The function $c$ combines two different lower bounds on the competitive ratio. With $c\equiv0$, $\alpha_\Lambda(g,0)$ is the natural weighted generalization of $\alpha(g)$. It does not improve over $0.5$ for arbitrary weights, since the first term of \eqref{eq:alphaw} is $\int_0^1\min\{g,1-g\}\,dx\le0.5$. The second lower bound corresponds to $c\equiv g$. The tradeoff function combines the two bounds better than a simple convex combination would (shown in Section~\ref{sec:weighted}), and we optimize it to find the best guarantee.

Theorems~\ref{thm:main} and~\ref{thm:weighted} hold for every continuous nondecreasing $g$ with $g(1)<1$. We evaluate them for the mixture family of Section~\ref{sec:model} in Appendix~\ref{app:numerics}. Remarkably, in AI-assisted numerical optimization over more general piecewise functions, the guarantees improved only in the third decimal place (by about $0.001$), so we omit those details. 
We note that the integrands of \eqref{eq:alphawL} are nondecreasing in $\sigma_\Lambda$, so the guarantee decreases as $\Lambda$ increases, converging to $\alpha_\infty(g,c)$ for arbitrary weights (see Table~\ref{tab:lambda} in Section~\ref{sec:results}).   

\subsection{Dual Certificate and Reduction to Per-Seed Inequality}\label{sec:dual}

\paragraph{Coupling.} In our analysis, \PG\ and $\opt$ use the same thresholds $\bz$. This ensures that if \PG\ probes $i$ at $t$, then it succeeds whenever $\opt$ succeeds on $(i,t)$. In addition, \PG\ uses the seeds $\bu$. In the analysis we consider one resource $i$ at a time. Let $\bu_{-i}$ denote the vector of seeds excluding $i$. With seed $u_i=u$, we let $\alg$ denote the run of \PG\ on the instance with seeds $\bu=(u,\bu_{-i})$ and thresholds $\bz$. We write $\algmi$ for the run of \PG\ on the instance with $i$ deleted, using the seeds $\bu_{-i}$ and the same thresholds $\bz$. Overloading notation, $\E[\PG]$ denotes the expected total reward of \PG. 

\paragraph{Candidate Dual Solution.} Initialise all duals to $0$. If \PG\ probes $(i,t)$, set
\[
\lambda_t:=r_i(1-g(u_i))\,\one\{z_t\le p_{it}\},\qquad \theta_i\ \mathrel{+}=\ r_ig(u_i)\,\one\{z_t\le p_{it}\},
\]
where $\mathrel{+}=$ means that the amount is added to the current value of $\theta_i$. Each match of $i$ contributes exactly $r_i$, so $\sum_t\lambda_t+\sum_i\theta_i$ equals the value of \PG\ on every sample path. 
Following the LP-free approach of \cite{GIU}, we compare these duals directly with \opt, one resource at a time. Let $O_i$ denote the set of arrivals at which $i$ is probed by $\opt$.

\begin{lemma}\label{lem:perres}
If for every $i\in U$
\begin{equation}\label{eq:perres}
\E\Big[\sum_{t\in O_i}\lambda_t\Big]+\E[\theta_i]\ \ge\ \alpha\,r_i\Pr[\opt\text{ matches } i],
\end{equation}
then $\E[\PG]\ge\alpha\,\E[\opt]$.
\end{lemma}
\begin{proof}
Each arrival lies in at most one set $O_i$, and $\lambda\ge0$. Hence $\E[\PG]=\E[\sum_t\lambda_t+\sum_i\theta_i]\ge\sum_i(\E[\sum_{t\in O_i}\lambda_t]+\E[\theta_i])\ge\alpha\sum_ir_i\Pr[\opt\text{ matches }i]=\alpha\,\E[\opt]$.
\end{proof}

\paragraph{Per-Seed Inequality.} To establish \eqref{eq:perres} with $\alpha>0.5$, we hereafter fix an arbitrary resource $i\in U$ and seeds $\bu_{-i}$, and write $u:=u_i$. Fixing $\bu_{-i}$ gives a stronger (conditional expectation) version of \eqref{eq:perres}. We will prove this stronger version by showing that at each arrival $t$ where $\opt$ could match $i$, the candidate dual solution ``pays'' an amount $r_ip_{it}\gamma(u,\pi_t)$, for an appropriately defined \emph{target} function $\gamma(x,\pi)\ge0$ of $x\in[0,1]$ and $\pi\ge0$. Here $\pi_t\ge0$ is a number determined by $\bu_{-i}$ and $\bz_{<t}$, which we specify later. In particular, $\pi_t$ is independent of $u$. 

\begin{lemma}\label{lem:perseed}
Suppose that 
for every $i$, every $u$ and every $\bu_{-i}$,
\begin{equation}\label{eq:perseed}
\E\Big[\sum_{t\in O_i}\lambda_t+\theta_i\ \Big|\ u_i=u,\ \bu_{-i}\Big]\ \ge\ r_i\,\E\Big[\sum_{t\in O_i}\one\{z_t\le p_{it}\}\,\gamma(u,\pi_t)\ \Big|\ u_i=u,\ \bu_{-i}\Big],
\end{equation}
where the expectations are over $\bz$ and $\pi_t$ is independent of $u$. If, in addition,
\[
\int_0^1\gamma(x,\pi)\,dx\ \ge\ \alpha \qquad \forall \pi\ge 0,
\]
then $\E[\PG]\ge\alpha\,\E[\opt]$.
\end{lemma}
\begin{proof}
Fix $\bu_{-i}$. The set $O_i$, the events $\{z_t\le p_{it}\}$ and the numbers $\pi_t$ do not depend on the seed $u_i$, which is uniformly sampled from $[0,1]$ and independent of all other randomness. Integrating \eqref{eq:perseed} over $u_i=u\in[0,1]$ therefore gives
\[
\E\Big[\sum_{t\in O_i}\lambda_t+\theta_i\ \Big|\ \bu_{-i}\Big]\ \ge\ r_i\,\E\Big[\sum_{t\in O_i}\one\{z_t\le p_{it}\}\int_0^1\gamma(x,\pi_t)\,dx\ \Big|\ \bu_{-i}\Big]\ \ge\ \alpha\,r_i\,\E\Big[\sum_{t\in O_i}\one\{z_t\le p_{it}\}\ \Big|\ \bu_{-i}\Big].
\]
Since $\opt$ stops probing $i$ after its first match, $\sum_{t\in O_i}\one\{z_t\le p_{it}\}=\one\{\opt\text{ matches }i\}$. Averaging over $\bu_{-i}$ gives \eqref{eq:perres} and by Lemma \ref{lem:perres}, the claim follows.
\end{proof}
\paragraph{Targets.} We now define the targets. For $x\in[0,1]$ and $\pi\ge0$ let
\[
\beta^1(x,\pi):=g(x)\,\one\{1-g(x)>\pi\},\qquad\beta^2(x,\pi):=\big(\sigma(x)g(x)+1-g(x)-\pi\big)\,\one\{1-g(x)>\pi\},
\]
and, with weights,
\[
w(x,\pi):=\min\Big\{1,\ \frac{1-g(x)-\pi}{(1-g(x))(1-\sigma_\Lambda(x))}\Big\}\qquad(\text{read as $1$ if $\sigma_\Lambda(x)=1$}),
\]
\begin{align*}
\beta^1_\Lambda(x,\pi)&:=\big(g(x)-c(x)+c(x)\,w(x,\pi)\big)\,\one\{1-g(x)>\pi\},\\
\beta^2_\Lambda(x,\pi)&:=\big(1-g(x)-\pi+(g(x)-c(x))\,\sigma_\Lambda(x)+c(x)\,w(x,\pi)\big)\,\one\{1-g(x)>\pi\}.
\end{align*}
Our two targets are
\[
\gamma(x,\pi):=\pi+\min\{\beta^1(x,\pi),\beta^2(x,\pi)\},\qquad \gamma_\Lambda(x,\pi):=\pi+\min\{\beta^1_\Lambda(x,\pi),\beta^2_\Lambda(x,\pi)\}.
\]

These targets are obtained by compacting the unwieldy expressions \eqref{eq:alpha} and \eqref{eq:alphawL}. In both expressions, the difference between the two terms in the outer minimum is merely technical: when $g(1)<1$, there is no $y\in[0,1]$ for which $1-g(y)=0$, so we have a separate first term to capture this possibility. If we extend the domain of $g$ beyond $1$ 
so that at some $y=y_1>1$ we have $g(y_1)=1$, then the second term (with integration limit adjusted to $\min\{y,1\}$) reduces to the first when $y=y_1$. Thus, we can capture the competitive ratio guarantees with just a (modified) second term from each expression. 
The next lemma formalizes this and its proof is in Appendix~\ref{app:integrate}.

\begin{lemma}\label{lem:integrate}
For every $\pi\ge0$,
\[
\int_0^1\gamma(x,\pi)\,dx\ \ge\ \alpha(g)\qquad\text{and}\qquad\int_0^1\gamma_\Lambda(x,\pi)\,dx\ \ge\ \alpha_\Lambda(g,c).
\]
Both bounds are tight: $\alpha(g)$ and $\alpha_\Lambda(g,c)$ are the infima of the two integrals over $\pi\ge0$.
\end{lemma}

\section{The Unweighted Setting}\label{sec:analysis}

 At a high level, the analysis has the same two steps as for classic OBM: we lower bound $\lambda_t$ and $\theta_i$ to show that they together pay for the target $\gamma(u,\pi_t)$. We do so by understanding how adding $i$ changes the matching output by \PG. In other words, we compare $\algmi$ with $\alg$.

Generalizing the analysis of \cite{DJK13} for OBM to OMSR, \cite{GU23} showed that the reduced price at every arrival in $\alg$ is no less than in $\algmi$, because including $i$ can only enlarge the set of resources available at each arrival. Lemma \ref{lem:cont} repackages this known lower bound on $\lambda_t$ in our notation. For completeness, we prove it in Appendix~\ref{app:deferred}.

\paragraph{Notation.} In the unweighted setting, $r_j=1$ for every resource $j\in U$. Throughout this section $i$, $u=u_i$ and $\bu_{-i}$ are fixed, so all seeds $\bu$ are fixed. Accordingly, every expectation $\E$ and probability $\Pr$ in this section is over the thresholds $\bz$ only, i.e., it is conditional on $\bu$. 
The reduced price of $j$ at $t$ is $p_{jt}(1-g(u_j))$. Let $P^{-i}_t$ be the reduced price of $\algmi$'s probe at $t$ ($0$ if none). For every arrival $t$ with an edge to $i$ let
\[
\pi_t:=\frac{P^{-i}_t}{p_{it}},\qquad\delta_t(u):=1-g(u)-\pi_t .
\]
So $\pi_t$ is the normalized reduced price of $\algmi$'s selection at $t$ and $\delta_t(u)$ is the margin by which the normalized reduced price $1-g(u)$ of $i$ exceeds that of $\algmi$'s selection. If the runs $\alg$ and $\algmi$ agree (select the same actions) up to $t$, then \PG\ probes $i$ at $t$ if $\delta_t(u)>0$. (In case $\delta_t(u)=0$, the tie breaking rule determines if $i$ is probed at $t$.) We say that $i$ \emph{beats} that selection if $\delta_t(u)>0$, i.e., if the reduced price of $i$ is strictly higher. For $(i,t)\in E$ define the \emph{boost}
\[
e_t:=\frac{P_t-P^{-i}_t}{p_{it}}=\frac{P_t}{p_{it}}-\pi_t,
\]
which is the (normalized) amount by which $\alg$'s reduced price at $t$ exceeds that of $\algmi$. The normalization in the denominator is the same as that of $\pi_t$, so that $P_t=p_{it}(\pi_t+e_t)$. \cite{GU23} showed that the boost is always non-negative, i.e., $e_t\geq 0$.
\begin{lemma}\label{lem:cont}
	Pathwise, at every arrival $t$, the set of resources available in $\alg$ at $t$ contains the set available in $\algmi$ at $t$. Consequently $P_t\ge P^{-i}_t$ and $e_t\geq 0$ for all $t$.
\end{lemma}

The next lemma says that, in expectation, $\lambda_t$ pays for $\pi_t+e_t$ at every arrival at which $\opt$ matches $i$. Its proof is included in Appendix~\ref{app:deferred}.

\begin{lemma}\label{lem:floor}
	$\E\big[\sum_{t\in O_i}\lambda_t\big]=\E\big[\sum_{t\in O_i}\one\{z_t\le p_{it}\}(\pi_t+e_t)\big]$.
\end{lemma}
The target at $t$ is
\[
\gamma(u,\pi_t)=\pi_t+\min\{\beta^1(u,\pi_t),\ \beta^2(u,\pi_t)\}.
\]
The value $\pi_t=P^{-i}_t/p_{it}$ is (i) independent of $u$, satisfying the requirement in Lemma~\ref{lem:perseed}, and (ii) always paid for by $\lambda_t$ (Lemma~\ref{lem:floor}). The second property implies that the boost $e_t$ and $\theta_i$ must together pay for the remaining target, which is $\min\{\beta^1(u,\pi_t),\beta^2(u,\pi_t)\}$. This is the main technical novelty in our subsequent analysis for the unweighted setting, described in Sections~\ref{sec:struct} and~\ref{sec:relarr}. We start by narrowing down what $\theta_i$ needs to pay and then prove that it pays this amount in expectation. The last step is where we use the correlations in the correlated model.

\begin{remark}\label{rem:boost}
Classic analyses of \PG\ typically ignore the boost and $\theta_i$ alone covers the remaining target, which in case of OBM is simply \[g(u)\one\{\delta_t(u)>0\}.\] \cite{Udw25} (in the deterministic Adwords setting) uses that the boost at every arrival $t$ with an edge to $i$ is at least $e_t\ge1-g(u)-\pi_t=\delta_t(u)$ when $i$ is available at \emph{every} arrival. By contrast, we will show (and use) the fact that the boost is also sometimes (i.e., on some sample paths) available at arrivals where $i$ itself is unavailable (such as when it is matched to an earlier arrival). Overall, the boost is particularly helpful for covering the remaining target in the troublesome instances such as Example~\ref{ex:release}, where the release of $j$ reduces $\theta_i$. 
\end{remark}



\subsection{What We Need from \texorpdfstring{$\theta_i$}{theta\_i}}\label{sec:struct}

To show the per-seed inequality, we need $\E[\theta_i]$ to carry the part of the target that $\lambda_t$ does not pay for. By Lemma~\ref{lem:floor}, $\lambda_t$ pays for $\pi_t+e_t$, so we define the \emph{demand} $D_t$ at $t$ as the portion of the target remaining after we subtract the contribution from $\lambda_t$. The term $\pi_t$ cancels and we have,
\begin{align*}
D_t:=\big(\gamma(u,\pi_t)-(\pi_t+e_t)\big)^+&=\big(\pi_t+\min\{\beta^1(u,\pi_t),\beta^2(u,\pi_t)\}-\pi_t-e_t\big)^+\\
&=\big(\one\{\delta_t(u)>0\}\,\min\{g(u),\ \sigma(u)g(u)+\delta_t(u)\}-e_t\big)^+.
\end{align*}
So the demand depends on $\algmi$ only through $\delta_t(u)$. Arrivals with $\delta_t(u)\le0$ (same as $1-g(u)\le \pi_t$) are the ones where $\algmi$ selects a resource with reduced price at least as high as that of $i$. There is no demand at these arrivals ($D_t=0$), because the indicator vanishes and $e_t\ge0$  ($\lambda_t$ fully covers the target).

The next lemma shows that $D_t$ is at most $g(u)$, and at most $\sigma(u)g(u)$ when $\alg$ probes at $t$ a resource that is at least as expensive as $i$. 
\begin{lemma}\label{lem:bounds}
Pathwise, for every $t$ with $(i,t)\in E$:
\begin{enumerate}[label=(\Alph*)]
\item $D_t\le g(u)$;
\item if $\alg$ probes at $t$ a resource whose reduced price is at least $p_{it}(1-g(u))$, then $e_t\ge\delta_t(u)$ and $D_t\le\sigma(u)g(u)$.
\end{enumerate}
\end{lemma}
\begin{proof}
(A) holds since $\one\{\delta_t(u)>0\}\min\{g(u),\sigma(u)g(u)+\delta_t(u)\}\le g(u)$ and $e_t\ge0$. To see (B), observe that when the reduced price of $\alg$'s probe at $t$ satisfies
\[
P_t=p_{it}(\pi_t+e_t)\ \ge\ p_{it}(1-g(u)),
\]
then $e_t\ \ge\ 1-g(u)-\pi_t\ =\ \delta_t(u).$ 
If $\delta_t(u)\le0$ then $D_t=0$; otherwise \[D_t\le\big(\sigma(u)g(u)+\delta_t(u)-\delta_t(u)\big)^+=\sigma(u)g(u).\]
\end{proof}

In particular, part~(B) applies whenever $i$ is available in $\alg$ at $t$, because the reduced price $P_t$ is then at least $p_{it}(1-g(u))$, the reduced price of $i$.

Now, $\E[\theta_i]$ needs to be large enough to cover the expected demand at the arrival where $\opt$ successfully matches $i$ (if any). Since $D_t=0$ when $\delta_t(u)\le0$, and $\opt$ can match $i$ at $t$ only if $z_t\le p_{it}$, a nonzero demand can arise only at the arrivals at which $i$ beats the probe of $\algmi$ and the edge to $i$ is successful. We denote the set of arrivals at which $\opt$ could make a nonzero demand by
\[
T_i:=\{t:\ (i,t)\in E,\ \delta_t(u)>0,\ z_t\le p_{it}\}.
\] 

 We do not know the probing order of $\opt$, so to simplify the comparison we look at the largest possible demand. Let $M$ be the largest demand at an arrival whose edge to $i$ is successful,
\[
M:=\max\{D_t:\ (i,t)\in E,\ z_t\le p_{it}\}=\max_{t\in T_i}D_t\qquad(\max\emptyset:=0).
\]
 Since the seeds are fixed, $M$ and $T_i$ are functions of $\bz$ alone; in particular they do not involve $\opt$. The next lemma shows that it suffices to compare $\theta_i$ with $M$, which greatly simplifies the subsequent analysis.

\begin{lemma}\label{lem:red}
	If $\E[\theta_i]\ge\E[M]$, then the per-seed inequality \eqref{eq:perseed} holds.
\end{lemma}
\begin{proof}
	By Lemma~\ref{lem:floor}, \eqref{eq:perseed} is equivalent to
	\[
	\E[\theta_i]\ \ge\ \E\Big[\sum_{t\in O_i}\one\{z_t\le p_{it}\}\big(\gamma(u,\pi_t)-\pi_t-e_t\big)\Big].
	\]
	There is at most one $t\in O_i$ with $z_t\le p_{it}$, because $\opt$ consumes $i$ at its first success on $i$. Hence, pathwise,
	\[
	\sum_{t\in O_i}\one\{z_t\le p_{it}\}\big(\gamma(u,\pi_t)-\pi_t-e_t\big)\ \le\ \sum_{t\in O_i}\one\{z_t\le p_{it}\}\,D_t\ \le\ M,
	\]
	and the hypothesis $\E[\theta_i]\ge\E[M]$ gives the inequality above.
\end{proof}

\subsection{An ``Almost'' Pathwise Proof of \texorpdfstring{$\E[\theta_i]\ge\E[M]$}{E[theta\_i] >= E[M]}}\label{sec:relarr}
On a sample path where $\alg$ matches $i$ we have $\theta_i=g(u)$, and $g(u)\ge M$ by Lemma~\ref{lem:bounds}(A). On paths where $\alg$ never matches $i$, $\theta_i=0$, but it is possible that $M>0$. This can only occur if some arrival $t$ with $\delta_t(u)>0$ has a successful edge $(i,t)\in E$, i.e., $z_t\le p_{it}$, but $\alg$ probes a resource other than $i$ at $t$. Example~\ref{ex:release} gives one such instance, where $j$ is released at arrival~$1$ and then displaces $i$ at arrival~$2$. In fact, every such sample path requires a release, a notion we make precise now.

\begin{definition}[Best Alternative and Releases]\label{def:release}
If $\alg$ probes $i$ at $\tau$, its \emph{best alternative} $J=J_\tau$ is the resource $\alg$ would probe at $\tau$ if $i$ were removed from its current available set (if any). We say $\alg$ \emph{releases} at $\tau$ if it unsuccessfully probes $i$ although a probe of $J$ would have succeeded. Formally, the best alternative $J$ is released at $\tau$ if $\alg$ probes $i$ at $\tau$, $J$ exists, and $p_{i\tau}<z_\tau\le p_{J\tau}$.
\end{definition}

In Example~\ref{ex:release}, $\alg$ releases its best alternative $J=j$ at arrival~$1$ whenever $1/4<z_1\le1/2$. A release at an arrival $\tau$ may also free a resource $J$ that never competes with or displaces $i$ at later arrivals. However, we show that a release has to occur on every troublesome path where $\theta_i=0$ but $M>0$. This is formalized in the next lemma, which says that, as long as a run of \PG\ that holds $i$ neither matches $i$ nor releases, removing $i$ changes the run only at the arrivals where it probes $i$. An immediate corollary is that $M=0$ without a release.

\begin{lemma}\label{lem:shadow}
	Consider two runs of \PG\ from some arrival $s$ onward, with the same seeds and thresholds, starting from available sets $S\cup\{i\}$ (run~1) and $S$ (run~2), where $i\notin S$. Suppose run~1 neither matches $i$ nor releases at any arrival $t\ge s$. Then at every arrival $t\ge s$ the two runs have the same available resources other than $i$. Moreover, at every arrival where run~1 does not probe $i$, both runs probe the same resource with the same outcome.
\end{lemma}
\begin{proof}
	We prove this by induction on $t\ge s$. The claim holds at $t=s$, where the available sets are $S\cup\{i\}$ and $S$. Suppose that just before arrival $t$ the available sets of the two runs differ only by $i$: let $S_t$ denote the available set in run~2 and $S_t\cup\{i\}$ the set available in run~1. It suffices to show that once $t$ departs, the available sets are again of the form $S_{t+1}\cup\{i\}$ and $S_{t+1}$, in run~1 and run~2 respectively, and that the two runs make the same probe with the same outcome if run~1 does not probe $i$ at $t$.
	
	If run~1 probes some $k\ne i$ at $t$, then $k$ is the resource selected from $S_t\cup\{i\}$ and $k\in S_t$, so by deterministic tie breaking run~2 probes $k$ as well, and both see the outcome $\one\{z_t\le p_{kt}\}$. If run~1 probes nothing, then $S_t$ has no neighbour of $t$ and run~2 probes nothing. If run~1 probes $i$, then $i$ fails ($z_t>p_{it}$) because run~1 never matches $i$ by assumption. Run~2 probes the best alternative $J$, if it exists. Since run~1 does not release at $t$, we have $z_t>p_{Jt}$, so the probe of $J$ fails in run~2. In all cases, the sets of available resources of the two runs again differ only by $i$ once $t$ departs.
\end{proof}

\begin{corollary}\label{cor:norelease}
	If $\alg$ never matches $i$ and never releases, then $T_i=\emptyset$ and $M=0$.
\end{corollary}
\begin{proof}
	We argue by contradiction. Suppose that $\alg$ never matches $i$ and never releases, but some arrival $t\in T_i$ exists, i.e.\ $(i,t)\in E$, $\delta_t(u)>0$ and $z_t\le p_{it}$. Apply Lemma~\ref{lem:shadow} from $s=1$ with $S=U\setminus\{i\}$; run~1 is then $\alg$ and run~2 is $\algmi$. Resource $i$ is available at $t$ in $\alg$, and its reduced price $p_{it}(1-g(u))$ strictly exceeds $P^{-i}_t$, because $\delta_t(u)>0$. By Lemma~\ref{lem:shadow}, $P^{-i}_t$ is the largest reduced price among the resources other than $i$ that are available in $\alg$ at $t$. So $\alg$ must probe $i$ at $t$, and the probe succeeds since $z_t\le p_{it}$, a contradiction. Finally, observe that $T_i=\emptyset$ implies $M=0$.
\end{proof}

So $\theta_i\ge M$ can fail only on paths where $\alg$ never matches $i$ but releases. We address these paths by shifting some value to them without changing $\E[\theta_i]$. The shifting is where edge correlations in the correlated model come in handy.

\paragraph{The Auxiliary Dual.} The dual of $i$ is a sum over the arrivals at which $\alg$ probes $i$:
\[
\theta_i=\sum_\tau\theta_{i\tau},\qquad \theta_{i\tau}:=g(u)\,\one\{\alg\text{ probes $i$ at $\tau$ and }z_\tau\le p_{i\tau}\},
\]
and at most one term is nonzero. To show $\E[\theta_i]\ge\E[M]$, we compare $M$ \emph{pathwise} with an auxiliary dual $\hat\theta_i=\sum_\tau\hat\theta_{i\tau}$ that has the same expectation as $\theta_i$ but is spread more evenly over the sample paths. We define $\hat\theta_{i\tau}$ by replacing $\theta_{i\tau}$ with its average over a set of values of $z_\tau$, at carefully selected arrivals $\tau$. We start by defining this set of arrivals.

\paragraph{Release Arrivals.} For an arrival $\tau$ write $\bz^{(\tau)}:=(\bz_{<\tau},0,\bz_{>\tau})$ for the threshold vector with the $\tau$-th threshold replaced by $0$. If $\alg$ probes $i$ at $\tau$, an event determined by $\bz_{<\tau}$, then the probe succeeds on $\bz^{(\tau)}$. Moreover, after $\tau$ the run of $\alg$ depends on $z_\tau$ only through the outcome of its probe at $\tau$, so its run on $\bz^{(\tau)}$ coincides with its run on every threshold vector that has the $\tau$-th threshold at most $p_{i\tau}$ and agrees with $\bz$ everywhere else. So we call this the run of $\alg$ \emph{in which $(i,\tau)$ succeeds}.
\begin{definition}[Release Arrival]\label{def:relarr}
An arrival $\tau$ is a \emph{release arrival} if
\begin{enumerate}[label=(\roman*)]
\item $\alg$ probes $i$ at $\tau$, a best alternative $J=J_\tau$ exists, and $p_{i \tau}<p_{J\tau}$, and
\item in the run of $\alg$ in which $(i,\tau)$ succeeds, at every arrival $t>\tau$ with $(i,t)\in E$ and $z_t\le p_{it}$, $\alg$ probes a resource whose reduced price is at least $p_{it}(1-g(u))$.
\end{enumerate}
\end{definition}

At a release arrival the conclusion of condition~(ii) holds whether or not the probe of $i$ succeeds. If the probe fails, $\alg$ continues after $\tau$ from the available set $S\cup\{i\}$, where $S$ is the set from which it continues if the probe succeeds, and it sees the same thresholds $\bz_{>\tau}$. The argument of Lemma~\ref{lem:cont}, applied to these two runs from arrival $\tau+1$ onward, shows that the run that holds $i$ has a reduced price at least as high at every later arrival. So at all later arrivals where the edge to $i$ is successful, $\alg$ probes resources that are at least as expensive as $i$, and, by Lemma~\ref{lem:bounds}(B), every later demand is at most $\sigma(u)g(u)$. This is the extra boost we alluded to in Remark \ref{rem:boost}. 

The next lemma shows that whether or not $\tau$ is a release arrival does not depend on $z_{\tau}$, which will be crucial for defining the auxiliary dual.

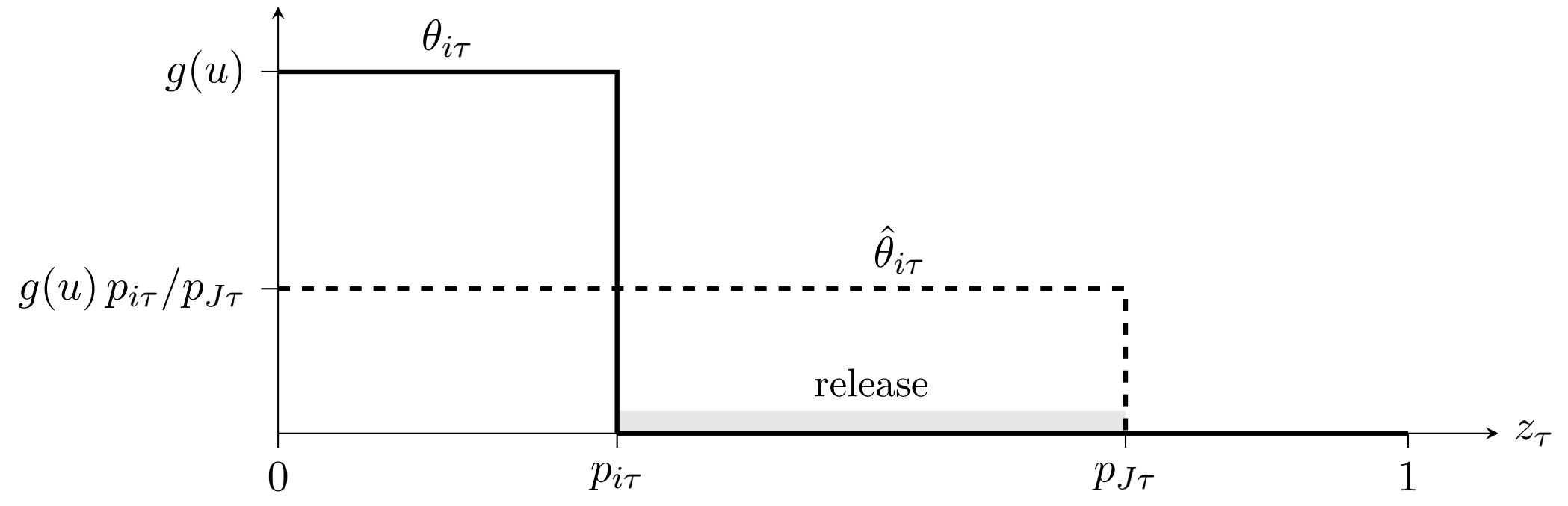
\begin{figure}[htbp]
\centering
\begin{tikzpicture}[x=10cm,y=3.2cm,>=stealth]
\fill[gray!20] (0.30,0) rectangle (0.75,0.06);
\draw[->] (0,0) -- (1.08,0) node[right] {$z_\tau$};
\draw[->] (0,0) -- (0,1.18);
\draw[very thick] (0,1) -- (0.30,1) -- (0.30,0);
\draw[very thick] (0.30,0) -- (1,0);
\draw[very thick,dashed] (0,0.4) -- (0.75,0.4) -- (0.75,0);
\foreach \x/\l in {0/$0$,0.30/$p_{i\tau}$,0.75/$p_{J\tau}$,1/$1$} \draw (\x,0) -- (\x,-0.04) node[below] {\l};
\draw (0,1) -- (-0.015,1) node[left] {$g(u)$};
\draw (0,0.4) -- (-0.015,0.4) node[left] {$g(u)\,p_{i\tau}/p_{J\tau}$};
\node[above] at (0.15,1) {$\theta_{i\tau}$};
\node[above] at (0.55,0.4) {$\hat\theta_{i\tau}$};
\node[above] at (0.525,0.06) {\small release};
\end{tikzpicture}
\caption{Averaging at a release arrival $\tau$, as a function of $z_\tau$ for fixed $\bz_{-\tau}$. The dual $\theta_{i\tau}$ (solid) equals $g(u)$ when the probe of $i$ succeeds ($z_\tau\le p_{i\tau}$) and $0$ otherwise; when $p_{i\tau}<z_\tau\le p_{J\tau}$ (shaded), $\alg$ releases $J$. The auxiliary dual $\hat\theta_{i\tau}$ (dashed) spreads the same total $g(u)p_{i\tau}$ evenly over $z_\tau\le p_{J\tau}$. By the crossing inequality \eqref{eq:cross}, its value is at least $\sigma(u)g(u)$.}\label{fig:avg}
\end{figure}

\begin{lemma}\label{lem:meas}
Condition~(i) of Definition~\ref{def:relarr}, the resource $J_\tau$ and the probability $p_{J\tau}$ are determined by $\bz_{<\tau}$, and condition~(ii) is determined by $\bz_{-\tau}$. In particular, whether $\tau$ is a release arrival does not depend on $z_\tau$.
\end{lemma}
\begin{proof}
$\alg$'s available set before $\tau$, its probe at $\tau$ and $J_\tau$ are functions of $\bz_{<\tau}$. Condition~(ii) is a property of the run of $\alg$ on $\bz^{(\tau)}$ and of the thresholds $z_t$ with $t>\tau$, and $\bz^{(\tau)}$ depends only on $\bz_{-\tau}$.
\end{proof}
We now define the auxiliary dual. For every arrival $\tau$ define
\[
\hat\theta_{i\tau}:=\begin{cases}
g(u)\,p_{i\tau}/p_{J\tau} & \text{if $\tau$ is a release arrival and } z_\tau\le p_{J\tau},\\
0 & \text{if $\tau$ is a release arrival and } z_\tau>p_{J\tau},\\
\theta_{i\tau} & \text{otherwise,}
\end{cases}
\qquad\hat\theta_i:=\sum_\tau\hat\theta_{i\tau}.
\]
At a release arrival, the value $g(u)$ that $\theta_{i\tau}$ takes when $z_\tau\le p_{i\tau}$ is spread evenly over the larger interval of values $z_\tau\le p_{J\tau}$; when $z_\tau>p_{J\tau}$ both are $0$. At all other arrivals $\hat\theta_{i\tau}=\theta_{i\tau}$. Clearly $\hat\theta_{i\tau}\ge0$, and, as we show next, $\hat\theta_{i\tau}$ has the same expectation as $\theta_{i\tau}$. Figure~\ref{fig:avg} illustrates the averaging.

\begin{lemma}\label{lem:mean}
For every arrival $\tau$, $\E[\hat\theta_{i\tau}\mid\bz_{-\tau}]=\E[\theta_{i\tau}\mid\bz_{-\tau}]$. Consequently, $\E[\hat\theta_i]=\E[\theta_i]$.
\end{lemma}
\begin{proof}
Fix $\bz_{-\tau}$. By Lemma~\ref{lem:meas}, this determines whether $\tau$ is a release arrival and also determines $J_\tau$ and $p_{J\tau}$. The threshold $z_\tau$ is uniform on $[0,1]$ and independent of $\bz_{-\tau}$. If $\tau$ is not a release arrival, then $\hat\theta_{i\tau}=\theta_{i\tau}$. Otherwise, $\alg$ probes $i$ at $\tau$, so $\E[\theta_{i\tau}\mid\bz_{-\tau}]=g(u)\,p_{i\tau}$, and
\[
\E[\hat\theta_{i\tau}\mid\bz_{-\tau}]=g(u)\,\frac{p_{i\tau}}{p_{J\tau}}\,p_{J\tau}=g(u)\,p_{i\tau}.
\]
The identity $\E[\hat\theta_i]=\E[\theta_i]$ follows by taking expectations on both sides and summing over $\tau$.
\end{proof}

Finally, we prove Theorem~\ref{thm:main} by a pathwise comparison between $\hat\theta_i$ and $M$.

\begin{proof}[Proof of Theorem~\ref{thm:main}]
By Lemmas~\ref{lem:perseed} and~\ref{lem:integrate}, it suffices to show \eqref{eq:perseed} for the target $\gamma(u,\pi_t)$, and by Lemma~\ref{lem:red} and Lemma~\ref{lem:mean}, it suffices to show $\E[\hat\theta_i]\ge\E[M]$. We show that $\hat\theta_i\ge M$ pathwise.

We observe that the greedy part of the price ensures that the probability $p_{i\tau}$ of the resource selected at $\tau$ cannot be much lower than the probability $p_{J\tau}$ of the best alternative. If $\alg$ probes $i$ at $\tau$ with best alternative $J$, then $p_{i\tau}(1-g(u))\ge p_{J\tau}(1-g(u_J))\ge p_{J\tau}(1-g(1))$. Since $p_{J\tau}>0$, this gives the \emph{crossing} inequality
\begin{equation}\label{eq:cross}
\frac{p_{i\tau}}{p_{J\tau}}\ \ge\ \frac{1-g(1)}{1-g(u)}=\sigma(u).
\end{equation}

Since every $\hat\theta_{i\tau}$ is nonnegative, it suffices to find, on every sample path, one arrival $\tau$ where $\hat\theta_{i\tau}\ge M$. There are three cases.

\emph{(a) $\alg$ matches $i$ at an arrival $\tau$ that is not a release arrival.} Then $\hat\theta_{i\tau}=\theta_{i\tau}=g(u)\ge M$ by Lemma~\ref{lem:bounds}(A).

\emph{(b) $\alg$ matches $i$ at a release arrival $\tau$.} Then $z_\tau\le p_{i\tau}<p_{J\tau}$, so $\hat\theta_{i\tau}=g(u)p_{i\tau}/p_{J\tau}\ge\sigma(u)g(u)$ by \eqref{eq:cross}. Let $t$ be an arrival with $(i,t)\in E$ and $z_t\le p_{it}$. If $t\le\tau$, then $i$ is available at $t$. If $t>\tau$, then, because $\tau$ is a release arrival and $(i,\tau)$ succeeds on this path, condition~(ii) says that $\alg$ probes at $t$ a resource whose reduced price is at least $p_{it}(1-g(u))$. In both cases Lemma~\ref{lem:bounds}(B) gives $D_t\le\sigma(u)g(u)$. Hence $M\le\sigma(u)g(u)\le\hat\theta_{i\tau}$.

\emph{(c) $\alg$ never matches $i$.} Then $i$ is available at every arrival, so $D_t\le\sigma(u)g(u)$ for every $t$ by Lemma~\ref{lem:bounds}(B). If $\alg$ never releases, $M=0$ by Corollary~\ref{cor:norelease}. Otherwise let $\tau$ be the last arrival at which $\alg$ releases. We claim that $\tau$ is a release arrival, which will ensure that $\hat\theta_{i\tau}\geq \sigma(u)g(u)$. Condition~(i) holds at $\tau$, since $p_{i\tau}<z_\tau\le p_{J\tau}$. For condition~(ii), compare after $\tau$ the path $\bz$ (run~1) with the run on $\bz^{(\tau)}$, where the threshold at $\tau$ is set to $0$ (run~2). Both runs are identical before $\tau$, and at $\tau$ both probe $i$, which fails in run~1 and succeeds in run~2. Both runs see the same thresholds after $\tau$ and their available sets are $S\cup\{i\}$ and $S$, where $i\not\in S$. By assumption, run~1 never matches $i$ and does not release after $\tau$, so Lemma~\ref{lem:shadow} applies. Take $t>\tau$ with $(i,t)\in E$ and $z_t\le p_{it}$. Run~1 does not probe $i$ at $t$, since the probe would succeed. As $i$ is available, run~1 probes a resource $k\ne i$ whose reduced price is at least $p_{it}(1-g(u))$, and by Lemma~\ref{lem:shadow} run~2 probes the same $k$. This is condition~(ii). So $\tau$ is a release arrival with $z_\tau\le p_{J\tau}$, and $\hat\theta_{i\tau}=g(u)p_{i\tau}/p_{J\tau}\ge\sigma(u)g(u)\ge M$ by \eqref{eq:cross}.
\end{proof}

\subsection{Why a Direct Generalization Fails}\label{sec:Ip}
In the vertex-weighted setting, let $\Lambda:=\max_{j,k\in U}r_j/r_k$ be the maximum ratio of weights of the instance, as in Theorem~\ref{thm:weighted}. With non-uniform weights, the argument of Section~\ref{sec:relarr} generalizes with one important change: the crossing inequality \eqref{eq:cross} no longer gives a constant value lower bound on $p_{i\tau}/p_{J\tau}$. Let
\[
\kappa_j:=\frac{r_j(1-g(u_j))}{r_i(1-g(u))}
\]
denote the ratio of the reduced prices of $j$ and $i$, each normalized by its success probability. If $\alg$ probes $i$ at $\tau$ with best alternative $J$, then $r_ip_{i\tau}(1-g(u))\ge r_Jp_{J\tau}(1-g(u_J))$, and
\[
\frac{p_{i\tau}}{p_{J\tau}}\ \ge\ \frac{r_J(1-g(u_J))}{r_i(1-g(u))}=\kappa_J.
\]
A natural generalization of the proof, given in Appendix~\ref{app:IpA}, gives the bound
\[
\hat\theta_{i\tau}\ \ge\ \kappa_J\,r_ig(u)
\]
at every release arrival $\tau$ with $z_\tau\le p_{J\tau}$. Since $\kappa_J\ge(1-g(1))/(\Lambda(1-g(u)))$, the natural replacement for the function $\sigma(u)$ in the target is
\[
\sigma_\Lambda(u)=\frac{1-g(1)}{\Lambda(1-g(u))}.
\]
In the worst case ($\Lambda\to\infty$), this function is identically $0$. With $\sigma\equiv0$ in the target $\gamma$, the first term of \eqref{eq:alpha} becomes $\int_0^1\min\{g,1-g\}\,dx\le0.5$. Therefore, by itself, the generalized argument is insufficient to establish a nontrivial result in the weighted setting. In the next section, we introduce a new lower bound that patches up the overall target precisely when $\Lambda$ is large. A combination of the two lower bounds gives our main result for the general setting.

\section{Vertex Weights}\label{sec:weighted}

We start with an instance on which the lower bound of Section~\ref{sec:analysis} is insufficient; the new lower bound of this section emerges from it. In the instance, several resources of low weight and high success probability compete with the resource $i$, which has a higher weight but a lower success probability at every arrival. Resource $i$ is probed at the first arrival, and this probe starts a chain of releases that displace $i$ at every later arrival. Consequently $\alg$ makes a single attempt to match $i$, whereas $\opt$ may probe $i$ at every arrival.

\begin{figure}[htbp]
\centering
\begin{tikzpicture}[x=1cm,y=1cm,
  arr/.style={draw,rectangle,minimum size=6mm,inner sep=1pt},
  res/.style={draw,circle,minimum size=7mm,inner sep=0pt},
  pg/.style={very thick},
  pgmi/.style={thick,dashed},
  other/.style={thick,dotted,black!60},
  lab/.style={fill=white,inner sep=1pt,font=\footnotesize,text=black}]
\foreach \t in {1,...,4} \node[arr] (a\t) at ({2.2*(\t-1)},0) {$\t$};
\node (a5) at (8.8,0) {$\cdots$};
\foreach \t in {1,...,4} \node[res] (j\t) at ({2.2*(\t-1)+1.1},1.7) {$j_{\t}$};
\node[res] (i) at (3.3,-2.1) {$i$};
\foreach \t in {1,...,4} \draw[pgmi] (j\t) -- (a\t);
\draw[pg] (i) -- node[lab,pos=0.55] {$p_1$} (a1);
\draw[pg] (j1) -- (a2);
\draw[pg] (j2) -- (a3);
\draw[pg] (j3) -- (a4);
\draw[pg] (j4) -- (a5);
\draw[other] (i) -- node[lab,pos=0.5] {$p_2$} (a2);
\draw[other] (i) -- node[lab,pos=0.5] {$p_3$} (a3);
\draw[other] (i) -- node[lab,pos=0.5] {$p_4$} (a4);
\begin{scope}[shift={(9.9,0.9)}]
\draw[pg] (0,0.6) -- (0.8,0.6) node[right,font=\small] {probe of \alg};
\draw[pgmi] (0,0) -- (0.8,0) node[right,font=\small] {probe of \algmi};
\draw[other] (0,-0.6) -- (0.8,-0.6) node[right,font=\small,text=black] {edge of $i$ that only \opt\ probes};
\end{scope}
\end{tikzpicture}
\caption{The release chain of Example~\ref{ex:chain}, first four arrivals. Squares are arrivals and circles are resources. Edges to $j_1,j_2,\dots$ have probability $1$, and the label of an edge of $i$ is its probability. $\algmi$ matches $j_t$ at every arrival $t<T$. $\alg$ probes $i$ at arrival $1$; at every later arrival $t$ it matches $j_{t-1}$, which $\algmi$ has used at arrival $t-1$ and which displaces $i$. $\opt$ probes $i$ at every arrival until it succeeds.}\label{fig:chain}
\end{figure}
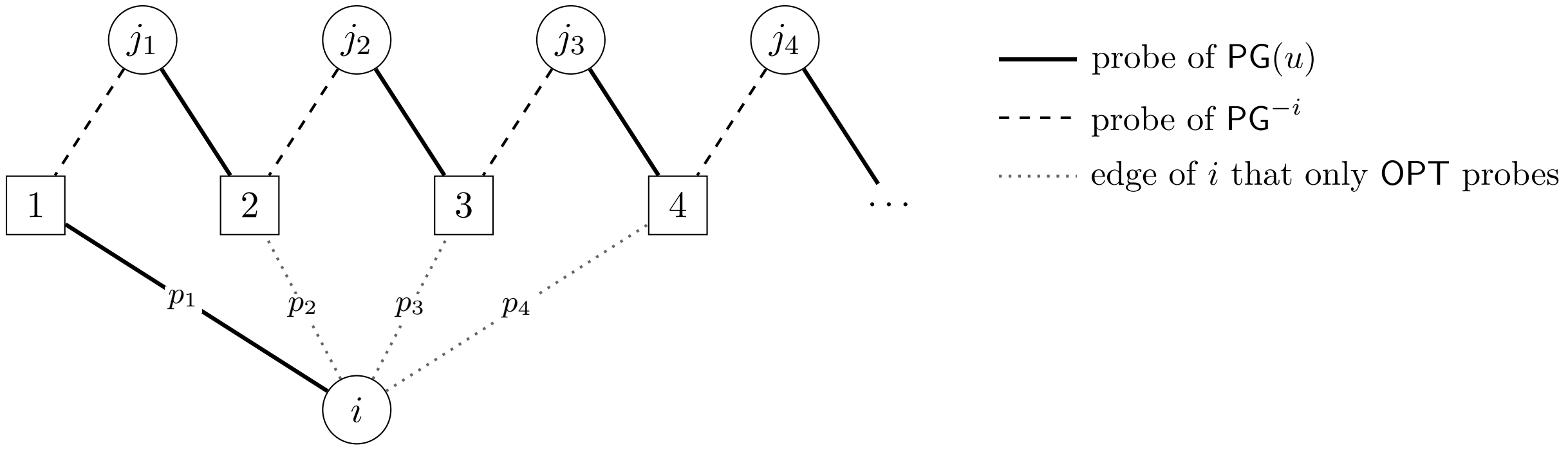
\begin{example}[A Release Chain]\label{ex:chain}
Let $r_i=1$ and fix the seed $u$ of $i$. There are $T$ arrivals, and $p_t$ denotes the (tiny) success probability of the edge $(i,t)$ for $t\in[T]$. Besides $i$ there are $T-1$ resources $j_1,\dots,j_{T-1}$; resource $j_t$ has edges to arrivals $t$ and $t+1$, and all edges to $j_1,\dots,j_{T-1}$ have probability $1$ (Figure~\ref{fig:chain}). Compared to $i$, these resources have tiny weight, proportional to the probabilities $p_t$. Specifically, weights and seeds are chosen so that $r_{j_t}(1-g(u_{j_t}))=\eta_t\,p_t(1-g(u))$ for some $\eta_t\in(0,1)$; this ensures that $i$ beats $j_t$ at arrival $t$. The probabilities $p_2,\dots,p_T$ are chosen small enough that \PG\ prefers $j_{t-1}$ to $i$ at arrival $t$.
\begin{itemize}[leftmargin=*]
\item \emph{The run $\algmi$:} At every arrival $t<T$, $\algmi$ probes $j_t$ and matches it; $j_t$ is the only available neighbour of $t$, because $j_{t-1}$ was matched at $t-1$. At arrival $T$, $\algmi$ probes nothing, and we set $\eta_T:=0$.
\item \emph{The run $\alg$:} At arrival $1$, $i$ beats $j_1$, so $\alg$ probes $i$ and matches it with probability $p_1$. Whatever the outcome, $j_1$ is available at arrival $2$, and since $p_2$ is small enough, $\alg$ prefers $j_1$ to both $i$ and $j_2$. So $\alg$ matches $j_1$, and $j_2$ remains available. In the same way, at every arrival $t\ge2$, $\alg$ matches $j_{t-1}$ and, if $t<T$, leaves $j_t$ for arrival $t+1$. Although $i$ beats $j_t$ at every $t$, the probe of $i$ at arrival $1$ starts a chain of releases that displaces $i$ at every $t\ge2$, and $\alg$ matches $i$ with probability $p_1$.
\item \emph{The benchmark:} $\opt$ probes $i$ at arrivals $1,2,\dots,T$ until the first success. It matches $i$ with probability $1-\prod_{t=1}^T(1-p_t)$.
\end{itemize}
\end{example}

The ratio $p_1/(1-\prod_t(1-p_t))$ of the probabilities with which $\alg$ and $\opt$ match $i$ can be made arbitrarily small. However, all is not lost because in this chain the crossing inequalities force the probabilities $p_t$ to shrink geometrically. We make this precise and use it to define a second lower bound on the target.

\emph{The success probabilities of $i$ shrink geometrically:} At arrival $t-1$, $\algmi$ probes $j_{t-1}$, and $\alg$ prefers $j_{t-1}$ to $i$ at arrival $t$. Both edges of $j_{t-1}$ have probability $1$, so its reduced price is the same at the two arrivals. Hence
\[
p_{t}(1-g(u))\ \le\ r_{j_{t-1}}(1-g(u_{j_{t-1}}))\ =\ \eta_{t-1}\,p_{t-1}(1-g(u)),\qquad\text{i.e.}\qquad p_{t}\ \le\ \eta_{t-1}\,p_{t-1}.
\]
So the probability that $\opt$'s edge to $i$ succeeds at $t$ is at most $p_1\prod_{s<t}\eta_s$.

\emph{The target of Section~\ref{sec:Ip} is not useful:} With arbitrary weights, the target $\gamma$ holds only with $\sigma_\infty\equiv0$ (Section~\ref{sec:Ip}). At every arrival of the chain, $\alg$ probes $i$ or $j_{t-1}$, whose reduced price is at least that of $i$, so the boost is at least $\delta_t(u)$. The target asks for nothing beyond the boost and this leaves the value $g(u)p_1$ that $\theta_i$ carries unused. On the other hand, the unweighted target, with $\sigma(u)=\frac{1-g(1)}{1-g(u)}$, cannot hold in this instance. Suppose $\alg$ breaks ties against $i$, and take $p_t=p_1\prod_{s<t}\eta_s$ for $t<T$. When $\eta_t$ is close to $1$ for every $t<T$, the unweighted target asks, in addition to the boost, for about $\sigma(u)g(u)$ at each arrival at which $\opt$'s edge to $i$ succeeds, in total about $\sigma(u)g(u)\big(1-\prod_t(1-p_t)\big)$, which can easily exceed $\E[\theta_i]=g(u)p_1$.

\emph{Splitting the gain of arrival $1$:} Imagine that the gain $g(u)p_1$ that $\alg$ makes at arrival $1$ is split across all arrivals, with arrival $t$ receiving the share $g(u)(1-\eta_t)$ whenever $\opt$'s edge to $i$ succeeds there (so arrivals at which the reduced price of $i$ is relatively higher than that of the resource matched to the arrival in the run without $i$ receive a larger share). By the geometric shrinkage, the shares add up to at most
\[
g(u)\sum_{t=1}^{T}p_t(1-\eta_t)\ \le\ g(u)\,p_1\sum_{t=1}^{T}(1-\eta_t)\prod_{s<t}\eta_s\ =\ g(u)\,p_1\Big(1-\prod_{t=1}^{T}\eta_t\Big)\ =\ g(u)\,p_1,
\]
because the sum telescopes and $\eta_T=0$. This suggests a second lower bound for the release chain, in which the split gain pays $g(u)(1-\eta_t)$ (on top of $\pi_t$, which $\lambda_t$ pays for). By itself this bound is not useful either because the share $1-\eta_t$ may be close to $0$. We show that a combination of the two bounds gives a nontrivial target.

\paragraph{Notation.} Throughout this section, $\Lambda\in[1,\infty]$ is the maximum ratio of vertex weights and $c$ is a function with $0\le c\le g$. As in the unweighted setting, we fix $i$ and the seeds $\bu$, with $u_i=u$. Recall that $\kappa_j=r_j(1-g(u_j))/(r_i(1-g(u)))\ge\sigma_\Lambda(u)$ for every $j\ne i$. For convenience we also set $\kappa_i:=1$. With weights, \[\pi_t:=\frac{P^{-i}_t}{r_ip_{it}}\] is the reduced price of $\algmi$'s probe at $t$, normalized by $r_ip_{it}$, and $\delta_t(u):=1-g(u)-\pi_t$ as before. For every arrival $t$ with an edge to $i$ let
\[
\eta_t:=\frac{\pi_t}{1-g(u)}.
\]
This is the ratio of the reduced price of $\algmi$'s probe at $t$ to the reduced price $r_ip_{it}(1-g(u))$ of $i$ at $t$; it generalizes the geometric factor of Example~\ref{ex:chain}. Observe that,
\[\delta_t(u)=(1-g(u))(1-\eta_t).\]
 We have $\delta_t(u)>0$ iff $\eta_t<1$. Like $\pi_t$, $\eta_t$ is a function of $\bz_{<t}$ alone (the seeds being fixed).

\paragraph{The Combined Target.} The observations above suggest two lower bounds on what the duals can pay at an arrival $t$ with $\delta_t(u)>0$, beyond the reduced price $\pi_t$ of $\algmi$, which $\lambda_t$ always pays for.
\begin{itemize}[leftmargin=*]
\item \emph{The first kind} is the target of Section~\ref{sec:analysis} with $\sigma_\Lambda$ in place of $\sigma$ (Section~\ref{sec:Ip}). It exceeds $\pi_t$ by $\min\{g(u),\ \sigma_\Lambda(u)g(u)+\delta_t(u)\}$. With arbitrary weights $\sigma_\infty\equiv0$, and the excess is $\min\{g(u),\delta_t(u)\}$.
\item \emph{The second kind} follows the intuition of splitting the gain. It exceeds $\pi_t$ by $g(u)\,w_t(u)$, where
\[
w_t(u)=\min\Big\{1,\ \frac{1-\eta_t}{1-\sigma_\Lambda(u)}\Big\}\qquad(\text{read as $1$ if $\sigma_\Lambda(u)=1$})
\]
is shorthand for the function $w(u,\pi_t)$ defined in Section~\ref{sec:alg}. In general, $w_t(u)\geq 1-\eta_t$ but with arbitrary weights $w_t(u)=1-\eta_t$, matching the share of arrival $t$ in the release chain.
\end{itemize}
Neither bound beats $0.5$ by itself. A convex combination with weights $1-\nu$ and $\nu$ would exceed $\pi_t$ by $(1-\nu)\min\{g(u),\sigma_\Lambda(u)g(u)+\delta_t(u)\}+\nu g(u)\,w_t(u)$. Instead, we combine the two bounds through the function $c$: a match of $i$ uses the part $c(u)$ of its (normalized) gain $g(u)$ for the second kind, by splitting it along the chain of releases it starts, and keeps the rest $g(u)-c(u)$ for the first kind. The combined target exceeds $\pi_t$ by
\[
\min\{g(u)-c(u),\ (g(u)-c(u))\sigma_\Lambda(u)+\delta_t(u)\}+c(u)\,w_t(u) ,
\]
which gives us the target $\gamma_\Lambda(u,\pi_t)$ at $t$, stated previously in Section~\ref{sec:alg}. For $c=\nu g$ this dominates the convex combination, and we optimize $c$ ($\le g$) to get the best combination. We prove Theorem~\ref{thm:weighted} in the next two subsections: Section~\ref{sec:wstruct} reduces \eqref{eq:perseed} for the target $\gamma_\Lambda(u,\pi_t)$ to two inequalities, one for each kind of bound, and Section~\ref{sec:split} proves the second.

\subsection{What We Need from \texorpdfstring{$\theta_i$}{theta\_i}}\label{sec:wstruct}
As in Section~\ref{sec:struct}, the demand at $t$ is the portion of the target that $\lambda_t$ does not pay for. By Lemma~\ref{lem:wfloor} in Appendix~\ref{app:Ip}, $\lambda_t$ pays for $\pi_t+e_t$, which gives us
\begin{align*}
D_t&:=\big(\gamma_\Lambda(u,\pi_t)-(\pi_t+e_t)\big)^+\\
&=\Big(\one\{\delta_t(u)>0\}\big(\min\{g(u)-c(u),\ (g(u)-c(u))\sigma_\Lambda(u)+\delta_t(u)\}+c(u)\,w_t(u)\big)-e_t\Big)^+.
\end{align*}
Same as before, there is no demand at arrivals with $\delta_t(u)\le0$. At the other arrivals, the demand depends on whether $\alg$ probes a resource at least as expensive as $i$. For $(i,t)\in E$ let
\[
v_t(u):=\begin{cases}\sigma_\Lambda(u)&\text{if $\alg$'s probe at $t$ has reduced price at least }r_ip_{it}(1-g(u)),\\ 1&\text{otherwise (including when $\alg$ probes nothing).}\end{cases}
\]
So $v_t(u)=\sigma_\Lambda(u)$ whenever $i$ is available in $\alg$ at $t$, and $v_t(u)=1$ is possible only after $\alg$ has matched $i$. The next lemma is the counterpart of Lemma~\ref{lem:bounds} and its proof is in Appendix~\ref{app:Ipstruct}.

\begin{lemma}\label{lem:clinear}
Pathwise, for every $t$ with $(i,t)\in E$ and $\delta_t(u)>0$:
\begin{enumerate}[label=(\Alph*)]
\item $D_t\le g(u)-c(u)+c(u)\,w_t(u)$;
\item if $\alg$ probes at $t$ a resource whose reduced price is at least $r_ip_{it}(1-g(u))$, then $e_t\ge\delta_t(u)$ and $D_t\le(g(u)-c(u))\sigma_\Lambda(u)+c(u)\,w_t(u)$.
\end{enumerate}
Overall, $D_t\le(g(u)-c(u))\,v_t(u)+c(u)\,w_t(u)$.
\end{lemma}

Let $M:=\max\{D_t:\ (i,t)\in E,\ z_t\le p_{it}\}=\max_{t\in T_i}D_t$, where 
\[T_i=\{t:\ (i,t)\in E,\ \delta_t(u)>0,\ z_t\le p_{it}\}\] is the set of arrivals at which $\opt$ could make a nonzero demand. The next lemma is Lemma~\ref{lem:red} with weights and its proof is also in Appendix~\ref{app:Ipstruct}.

\begin{lemma}\label{lem:wred}
If $\E[M]\le \frac{1}{r_i}\E[\theta_i]=g(u)\Pr[\alg\text{ matches }i]$, then \eqref{eq:perseed} holds for the target $\gamma_\Lambda$.
\end{lemma}
Let
\[
V:=\max_{t\in T_i}v_t(u)\qquad\text{and}\qquad W:=\max_{t\in T_i}w_t(u)\qquad(\max\emptyset:=0)
\]
denote the two parts of $M$ that $\frac{1}{r_i}\theta_i$ has to supply. By Lemma~\ref{lem:clinear}, on every path
\begin{equation}\label{eq:Msplit}
M\ \le\ (g(u)-c(u))\,V\ +\ c(u)\,W .
\end{equation}
To prove Theorem~\ref{thm:weighted} we show that $\E[V]$ and $\E[W]$ are both at most $\Pr[\alg\text{ matches }i]$.

\begin{lemma}\label{prop:Ip} The following statements hold.
\begin{enumerate}[label=(\Alph*)]
\item $\E[V]\ \le\ \Pr[\alg\text{ matches }i]$.
\item $\E[W]\ \le\ \Pr[\alg\text{ matches }i]$.
\end{enumerate}
\end{lemma}

Part~(A) is the weighted version of the almost pathwise proof of Theorem~\ref{thm:main}. 
The proof is in Appendix~\ref{app:IpA}. Part~(B) is the new term in the target and it formalizes the split in the release chain. It says that, in expectation, a match of $i$ pays the (normalized) share $w_t(u)\ge1-\eta_t$ at every arrival at which $\opt$ could make a nonzero demand, including those at which $\alg$ does not probe $i$. We prove it in Section~\ref{sec:split}.

\begin{proof}[Proof of Theorem~\ref{thm:weighted}]
By Lemmas~\ref{lem:perseed}, \ref{lem:integrate} and~\ref{lem:wred}, it suffices to show that $\E[M]\le g(u)\Pr[\alg$ matches $i]$, which follows by taking expectations in \eqref{eq:Msplit} and applying Lemma~\ref{prop:Ip},
\[
\E[M]\ \le\ \big(g(u)-c(u)+c(u)\big)\Pr[\alg\text{ matches }i]\ =\ g(u)\Pr[\alg\text{ matches }i].\qedhere
\]
\end{proof}

\subsection{Proof of \texorpdfstring{Lemma~\ref{prop:Ip}(B)}{part (B)} (Generalized Gain Splitting)}\label{sec:split}


Let us first replace $\E[W]$ by an upper bound that is simpler to work with. Since $0\le w_t(u)\le1$, every factor $1-w_t(u)$ lies in $[0,1]$, so the product of these factors over $t\in T_i$ is no larger than any single factor. Hence, on every path,
\begin{equation}\label{eq:Wprod}
W=\max_{t\in T_i}w_t(u)\ \le\ 1-\prod_{t\in T_i}\big(1-w_t(u)\big),
\end{equation}
where the empty product is $1$. We will show that the expectation of the right side of \eqref{eq:Wprod} is at most $\Pr[\alg\text{ matches }i]$, which implies Lemma~\ref{prop:Ip}(B). The expression $1-\prod_{t\in T_i}\big(1-w_t(u)\big)$  involves every arrival of $T_i$, whereas $\alg$ may probe $i$ at only a subset of the arrivals with $\delta_t(u)>0$ (in fact, $i$ is probed exactly once in the release chain). From another perspective, we will show that it is always feasible to split the gain from matching $i$ in $\alg$ among all arrivals at which $\opt$ could make a demand, so that, in expectation, every arrival $t\in T_i$ gets the share $w_t(u)$. 
We first introduce additional notation and bound how much removing a single resource from the available set can change the probability that $i$ is never matched.   

\paragraph{Additional Notation.} We write $A^{-i}_t$ for the set of resources available in $\algmi$ at $t$. So $A^{-i}_1=U\setminus\{i\}$, where $U$ is the set of all resources, and if $\algmi$ probes $j$ at $t$, then $A^{-i}_{t+1}=A^{-i}_t\setminus\{j\}$ when $z_t\le p_{jt}$ and $A^{-i}_{t+1}=A^{-i}_t$ otherwise. For an arrival $t$ and a set $A$ of resources (and fixed seeds $\bu$), let $Q_t(A)$ be the probability, over $\bz_{\ge t}$, that the run of \PG\ that starts at arrival $t$ with available set $A$ never matches $i$. Let $Q_t(A):=0$ if $i\notin A$, and $Q_{T+1}(A):=\one\{i\in A\}$. Then $Q_t(A)=\E_{z_t}[Q_{t+1}(A')]$, where $A'$ is the available set after the probe at $t$, and
\[
Q_1(U)=\Pr[\alg\text{ never matches }i].
\]
By Lemma~\ref{lem:cont}, for $i\in A'\subseteq A$ we have $Q_t(A')\le Q_t(A)$, because a run that starts with fewer resources matches $i$ whenever a run with more resources does.

The next lemma bounds the loss in the other direction: removing a single resource $j$ from $A$ lowers the probability that $i$ is never matched by at most the factor $1-\kappa_j$. The intuition is as follows. Removing $j$ makes no difference until the first arrival $t$ at which the run from $A$ probes $j$. There the run from $A\setminus\{j\}$ probes the next best resource $\ell$ instead. If $\ell=i$, then, since the run from $A$ prefers $j$ to $i$, we have $p_{it}\le\kappa_jp_{jt}\le\kappa_j$. Therefore, the probe of $i$ caused by the removal succeeds with probability at most $\kappa_j$. If $\ell\ne i$, then $\kappa_\ell p_{\ell t}\le\kappa_jp_{jt}$, and the probe of $\ell$ lowers the probability that $i$ is never matched by at most the factor $1-\kappa_\ell p_{\ell t}\ge1-\kappa_jp_{jt}$. Using induction over arrivals shows that the overall loss is at most a factor $1-\kappa_j$. 

\begin{lemma}\label{lem:remove}
For every arrival $t$, every set $A$ of resources and every $j\in A\setminus\{i\}$,
\[
Q_t(A\setminus\{j\})\ \ge\ (1-\kappa_j)\,Q_t(A).
\]
\end{lemma}

\begin{proof}
We prove the claim by backward induction on $t$, simultaneously for all sets $A$ and all $j\in A\setminus\{i\}$. The claim is trivial if $i\notin A$, since then both sides are $0$. At $t=T+1$, it holds since $Q_{T+1}(A\setminus\{j\})=Q_{T+1}(A)=\one\{i\in A\}$. Let $t\le T$, and let $k$ be the resource that \PG\ probes at $t$ from $A$, i.e.\ the neighbour of $t$ in $A$ with the highest reduced price (if there is none, \PG\ probes nothing, and we treat this as $k\ne j$).

\emph{Case $k\ne j$.} We show a stronger statement conditional on every fixed value of $z_t$. Since $k$ lies in $A\setminus\{j\}$, by deterministic tie breaking \PG\ from $A\setminus\{j\}$ also probes $k$ at $t$, with the same outcome $\one\{z_t\le p_{kt}\}$. Let $A'$ and $A'\setminus\{j\}$ be the available sets after arrival $t$ of the runs from $A$ and $A\setminus\{j\}$, respectively. Note that $A'$ contains $j$, so by the induction hypothesis at $t+1$, for every value of $z_t$,
\[
Q_{t+1}(A'\setminus\{j\})\ \ge\ (1-\kappa_j)\,Q_{t+1}(A').
\]
Therefore, taking expectations over $z_t$,
\[
Q_t(A\setminus\{j\})\ =\ \E_{z_t}\big[Q_{t+1}(A'\setminus\{j\})\big]\ \ge\ (1-\kappa_j)\,\E_{z_t}\big[Q_{t+1}(A')\big]\ =\ (1-\kappa_j)\,Q_t(A).
\]

\emph{Case $k=j$.} Then \PG\ from $A\setminus\{j\}$ probes the best alternative $\ell$ of $j$ in $A\setminus\{j\}$ (with $p_{\ell t}:=0$ if there is none). Since $j$ has the highest reduced price in $A$, $r_\ell p_{\ell t}(1-g(u_\ell))\le r_jp_{jt}(1-g(u_j))$, i.e.,
\begin{equation}\label{eq:removeA}
\kappa_\ell\,p_{\ell t}\ \le\ \kappa_j\,p_{jt},
\end{equation}
which for $\ell=i$ is the inequality $p_{it}\le\kappa_jp_{jt}$. We compare both runs with the run from $A\setminus\{j\}$ at $t+1$.
\begin{enumerate}[leftmargin=*]
\item \emph{The run from $A$.} It matches $j$ with probability $p_{jt}$, so
\[
Q_t(A)=p_{jt}\,Q_{t+1}(A\setminus\{j\})+(1-p_{jt})\,Q_{t+1}(A).
\]
By the induction hypothesis, $(1-\kappa_j)\,Q_{t+1}(A)\le Q_{t+1}(A\setminus\{j\})$, hence
\[
(1-\kappa_j)\,Q_t(A)\ \le\ \big(p_{jt}(1-\kappa_j)+1-p_{jt}\big)\,Q_{t+1}(A\setminus\{j\})\ =\ (1-\kappa_jp_{jt})\,Q_{t+1}(A\setminus\{j\}).
\]
\item \emph{The run from $A\setminus\{j\}$.} It matches $\ell$ with probability $p_{\ell t}$, so
\begin{align*}
Q_t(A\setminus\{j\})&=p_{\ell t}\,Q_{t+1}(A\setminus\{j,\ell\})+(1-p_{\ell t})\,Q_{t+1}(A\setminus\{j\})\\
&\ge\big(p_{\ell t}(1-\kappa_\ell)+1-p_{\ell t}\big)\,Q_{t+1}(A\setminus\{j\})\ =\ (1-\kappa_\ell\,p_{\ell t})\,Q_{t+1}(A\setminus\{j\})\\
&\ge(1-\kappa_jp_{jt})\,Q_{t+1}(A\setminus\{j\}).
\end{align*}
The first inequality is the induction hypothesis for the removal of $\ell$ from $A\setminus\{j\}$ at $t+1$; for $\ell=i$ it holds because $Q_{t+1}(A\setminus\{j,i\})=0$ and $\kappa_i=1$. The second inequality is \eqref{eq:removeA}.
\end{enumerate}
Combining the two items,
\[
Q_t(A\setminus\{j\})\ \ge\ (1-\kappa_jp_{jt})\,Q_{t+1}(A\setminus\{j\})\ \ge\ (1-\kappa_j)\,Q_t(A).
\]
\end{proof}

\begin{proof}[Proof of Lemma~\ref{prop:Ip}(B)]
Taking expectations in \eqref{eq:Wprod} gives $\E[W]\le1-\E[\prod_{t\in T_i}(1-w_t(u))]$. So it suffices to show that
\begin{equation}\label{eq:product}
	\E\Big[\prod_{t\in T_i}\big(1-w_t(u)\big)\Big]\ \geq\ \Pr[\alg\text{ \emph{does not} match }i]\ =\ Q_1(U).
\end{equation}
 We prove this by induction on the number $T$ of arrivals, for all instances whose weights lie within a factor $\Lambda$ of each other. With no arrivals both sides equal $1$. Let $T\ge1$ and assume that \eqref{eq:product} holds for all such instances with $T-1$ arrivals. Let $j$ be $\algmi$'s probe at arrival $1$ (with $p_{j1}:=0$ if there is none), so that $A^{-i}_2=U\setminus\{i,j\}$ when $z_1\le p_{j1}$ and $A^{-i}_2=U\setminus\{i\}$ otherwise. We condition on the threshold $z_1$ of the first arrival (fix $z_1$).

We start by reframing the induction hypothesis. Arrivals $2,\dots,T$ with the resources $A^{-i}_2\cup\{i\}$ form an instance with $T-1$ arrivals whose weights again lie within a factor $\Lambda$ of each other. On this instance, the run of \PG\ without $i$ starts from $A^{-i}_2$, so it is simply the continuation of $\algmi$ after $t=1$. Hence, for $t\ge2$, the values $\eta_t$, $w_t(u)$, and the events $t\in T_i$, are the same for this instance as for the original instance. 
The run from $A^{-i}_2\cup\{i\}$ on this instance never matches $i$ with probability $Q_2(A^{-i}_2\cup\{i\})$. Since $z_2,\dots,z_T$ are independent of $z_1$, the hypothesis applied conditionally on $z_1$ gives 
\[
\E\Big[\prod_{t\in T_i,\,t\ge2}\big(1-w_t(u)\big)\ \Big|\ z_1\Big]\ \ge\ Q_2(A^{-i}_2\cup\{i\})\qquad \forall z_1\in[0,1].
\]
 Multiplying by the factor $1-w_1(u)\,\one\{1\in T_i\}\in[0,1]$ of arrival $1$, which is determined by $z_1$, and taking expectation over $z_1$,
\begin{equation}\label{eq:onearr}
\E\Big[\prod_{t\in T_i}\big(1-w_t(u)\big)\Big]\ \ge\ \E\Big[\big(1-w_1(u)\,\one\{1\in T_i\}\big)\,Q_2\big(A^{-i}_2\cup\{i\}\big)\Big].
\end{equation}

 It remains to show that the right side of \eqref{eq:onearr} is at least $Q_1(U)$. Recall that $1\in T_i$ iff $(i,1)\in E$, $\delta_1(u)>0$ (i.e., $\eta_1<1$) and $z_1\le p_{i1}$, and that $j$ is $\algmi$'s probe at arrival $1$. There are two cases.

\emph{Case $\alg$ does not probe $i$ at arrival $1$.} Then $\alg$ and $\algmi$ make the same probe at arrival $1$, or both probe nothing. 
The two probes have the same outcome, so the available set of $\alg$ after arrival $1$ is $A^{-i}_2\cup\{i\}$, and
\[
Q_1(U)\ =\ \E_{z_1}\big[Q_2(A^{-i}_2\cup\{i\})\big].
\]
Moreover, $1\notin T_i$, because either $(i,1)\notin E$, or $(i,1)\in E$ and the reduced price of $j$ at arrival $1$ is at least that of $i$, so that $\eta_1\ge1$. Hence, $1-w_1(u)\,\one\{1\in T_i\}=1$, and the right side of \eqref{eq:onearr} equals $Q_1(U)$.

\emph{Case $\alg$ probes $i$ at arrival $1$.} Then $p_{i1}>0$, and $\eta_1\le1$ because the reduced price of $i$ at arrival $1$ is at least that of $j$. We compute both $Q_1(U)$ and the right side of \eqref{eq:onearr} for comparison. 
\begin{itemize}[leftmargin=*]
\item \emph{The value $Q_1(U)$.} $\alg$ matches $i$ if $z_1\le p_{i1}$. Otherwise its probe fails, no resource is used at arrival $1$, and $\alg$ continues from $U$. Hence
\[
Q_1(U)\ =\ (1-p_{i1})\,Q_2(U).
\]
\item \emph{The right side of \eqref{eq:onearr}.} Since $1\in T_i$ implies $z_1\le p_{i1}$, we have
 \[1-w_1(u)\,\one\{1\in T_i\}\geq 1-w_1(u)\,\one\{z_1\le p_{i1}\}.\] Moreover, $A^{-i}_2\cup\{i\}=U\setminus\{j\}$ if $z_1\le p_{j1}$, and $A^{-i}_2\cup\{i\}=U$ otherwise. So the right side of \eqref{eq:onearr} is at least
\[
\E_{z_1}\Big[\big(1-w_1(u)\,\one\{z_1\le p_{i1}\}\big)\Big(\one\{z_1\le p_{j1}\}\,Q_2(U\setminus\{j\})+\one\{z_1>p_{j1}\}\,Q_2(U)\Big)\Big].
\]
\end{itemize}
There are two cases depending on whether $p_{j1}\leq p_{i1}$.

\emph{Subcase $p_{j1}\le p_{i1}$} (this includes the case that $\algmi$ probes nothing, where $p_{j1}=0$ and the first term below vanishes). The intervals $z_1\le p_{j1}$, $p_{j1}<z_1\le p_{i1}$ and $z_1>p_{i1}$ give the three terms of
\begin{align*}
&(1-w_1(u))\,p_{j1}\,Q_2(U\setminus\{j\})\ +\ (1-w_1(u))\,(p_{i1}-p_{j1})\,Q_2(U)\ +\ (1-p_{i1})\,Q_2(U)\\
&\qquad\ge\ (1-p_{i1})\,Q_2(U)\ =\ Q_1(U),
\end{align*}
where the inequality holds because $w_1(u)\le1$, so that the first two terms are nonnegative.

\emph{Subcase $p_{j1}>p_{i1}$.} The intervals $z_1\le p_{i1}$, $p_{i1}<z_1\le p_{j1}$ and $z_1>p_{j1}$ give the three terms of
\begin{align*}
&(1-w_1(u))\,p_{i1}\,Q_2(U\setminus\{j\})\ +\ (p_{j1}-p_{i1})\,Q_2(U\setminus\{j\})\ +\ (1-p_{j1})\,Q_2(U)\\
&\qquad=\ \big(p_{j1}-w_1(u)\,p_{i1}\big)\,Q_2(U\setminus\{j\})\ +\ (1-p_{j1})\,Q_2(U).
\end{align*}
Since $Q_1(U)=(1-p_{i1})\,Q_2(U)=(p_{j1}-p_{i1})\,Q_2(U)+(1-p_{j1})\,Q_2(U)$, the terms with $1-p_{j1}$ agree, and it remains to show that
\begin{equation}\label{eq:release1}
\big(p_{j1}-w_1(u)\,p_{i1}\big)\,Q_2(U\setminus\{j\})\ \ge\ (p_{j1}-p_{i1})\,Q_2(U).
\end{equation}
The right side is the probability that the probe of $i$ at arrival~$1$ fails while $\algmi$ matches $j$, so that $j$ is released and $i$ is never matched from $t=2$ on. 
By Lemma~\ref{lem:remove} (with $t=2$ and $A=U$), $Q_2(U\setminus\{j\})\ge(1-\kappa_j)\,Q_2(U)$. Therefore,
\begin{align*}
	\big(p_{j1}-w_1(u)\,p_{i1}\big)\,Q_2(U\setminus\{j\})&\ \ge\ \big(p_{j1}-w_1(u)\,p_{i1}\big)(1-\kappa_j)\,Q_2(U)\\
	&\ \ge\ \big(p_{j1}-\tfrac{1-\eta_1}{1-\sigma_{\Lambda}(u)}\,p_{i1}\big)(1-\kappa_j)\,Q_2(U)\\
	&\ \ge\ \big(p_{j1}(1-\kappa_j)-(1-\eta_1)\,p_{i1}\big)\,Q_2(U)\\
	&\ =\ (p_{j1}-p_{i1})\,Q_2(U).
\end{align*}
The second step follows from
\[
w_1(u)\ \le\ \frac{1-\eta_1}{1-\sigma_\Lambda(u)},
\]
since $1-\kappa_j>0$ (see below). In the third step we use $\sigma_\Lambda(u)\le\kappa_j$, so that $(1-\kappa_j)/(1-\sigma_\Lambda(u))\le1$. In the last step we use the definition of $\eta_1$: the reduced price of $j$ at arrival $1$ is $\eta_1$ times that of $i$, i.e., $r_jp_{j1}(1-g(u_j))=\eta_1\,r_ip_{i1}(1-g(u))$, so that
\[
\kappa_j\ =\ \eta_1\,\frac{p_{i1}}{p_{j1}}\ <\ 1,\qquad\text{i.e.,}\qquad p_{j1}\kappa_j=\eta_1p_{i1}.
\]
This shows \eqref{eq:release1}. In all cases the right side of \eqref{eq:onearr} is at least $Q_1(U)$, which completes the induction.
\end{proof}

\section{Conclusion and Open Problems}\label{sec:conclusion}
We showed that Perturbed Greedy beats $0.5$ for (vertex-weighted) OMSR with arbitrary edge probabilities against the adaptive offline benchmark. The perturbation functions we use mix Greedy and Ranking. For the best functions in the mixture family, \PG\ is $\HeadUT$-competitive without weights and $\HeadWT$-competitive with vertex weights. 
We close with three open directions.

\paragraph{The Competitive Ratio of Stochastic Rewards.} The best competitive ratio for OMSR with arbitrary probabilities is not known, with or without weights. While \PG\ is not better than $0.6318$-competitive for any $g$ in the vertex-weighted setting (Proposition~\ref{prop:noopt}), no upper bound below $1-1/e$ is known in the unweighted setting. Several interesting questions remain open. Is $1-1/e$ achievable, with or without weights? Is there a separation between the best competitive ratios for the weighted and unweighted settings? Does \PG\ achieve the best guarantee for general OMSR?

\paragraph{Beyond Stochastic Rewards.} \citet{Udw26} proposes a generalization of online submodular welfare maximization \citep{KPV13} that captures many well-studied forms of post-allocation uncertainty (including OMSR). In this generality, non-adaptive Greedy is $0.5$-competitive, and by a recent unconditional hardness result for online submodular welfare maximization \citep{UW26}, no algorithm does better. So in general, non-adaptivity is optimal and adaptivity does not help. A natural open question is to determine the most general setting in which adaptive algorithms can beat $0.5$ and outperform non-adaptive ones. Specifically, can $0.5$ be beaten for OMSR with patience, where several edges can be probed at each arrival \citep{BGMS25,BM25}, or for online assortments, where each arrival is offered a set of resources to (randomly) choose from \citep{GNR14}? \PG\ extends naturally to both models, but adding a resource can change the entire set of resources offered to an arrival, so even basic properties, such as the fact that adding a resource can only enlarge the set of available resources at every arrival (Lemma~\ref{lem:cont}), may not hold.

\paragraph{Reusable Resources.} \cite{GIU} point out a connection between stochastic rewards and the online allocation of reusable resources with stochastic usage durations. A failed probe is like a use of the resource whose duration is $0$, so the resource returns at once, and a successful probe is like a use whose duration is $+\infty$. In this view, the usage distribution of a resource depends on the arrival. In general, no algorithm has a constant competitive ratio when usage distributions are arrival dependent \citep{GGISUW22}, even when durations are deterministic \citep{GIU}. Given that OMSR admits a ratio better than $0.5$, it is natural to ask which other families of arrival-dependent usage distributions admit such a result. Very recently, \citet{BTW26} show that Ranking beats $0.5$ when the usage distributions are independent of arrivals. It would be interesting to understand if a combination of their approach and ours can answer this question.
\clearpage
\appendix
\addtocontents{toc}{\protect\contentsline{section}{Appendix}{}{}}
\begin{center}
{\LARGE\bfseries Appendix}
\end{center}
\counterwithin{theorem}{section}
\counterwithin{example}{section}
\counterwithin{remark}{section}
\section{Perturbation Functions and Their Guarantees}\label{app:numerics}
This appendix gives the perturbation functions behind \infthm{}(\ref{inf:unweighted}) and~(\ref{inf:weighted}) and Table~\ref{tab:lambda}, and states their guarantees formally. All of them belong to the mixture family \eqref{eq:sband},
\[
g(x)=A\,e^{k\,s(x)}+B,\qquad s(x)=\min\{x,\lo\}+\max\{0,x-\hi\},
\]
where $s$ is the effective seed of the band $[\lo,\hi]$ (Section~\ref{sec:model}). With weights we use the tradeoff function $c(x)=\nu g(x)$ for a constant $\nu\in[0,1]$. The parameters, listed in Table~\ref{tab:bands}, were optimized numerically with AI assistance, so the true optimal functions of the mixture family may give a slightly better competitive ratio lower bound. 

\begin{table}[htbp]
\centering\small
\begin{tabular}{lcccccc}
\toprule
weights & $[\lo,\hi]$ & $A$ & $k$ & $B$ & $\nu$ & guarantee $\ge$\\
\midrule
unweighted ($\Lambda=1$) & $[0.6457,1]$ & $0.0470$ & $3.1813$ & $0.3613$ & -- & $\HeadU$\\
within a factor $\Lambda=2$ & $[0.7143,1]$ & $0.0247$ & $3.3209$ & $0.4134$ & $0.0514$ & $0.5610$\\
within a factor $\Lambda=3$ & $[0.8399,1]$ & $0.1435$ & $1.4448$ & $0.2347$ & $0.4101$ & $0.5469$\\
within a factor $\Lambda=10$ & $[0.1214,0.4318]$ & $0.4060$ & $0.9332$ & $0.0026$ & $0.6262$ & $0.5358$\\
arbitrary ($\Lambda=\infty$) & $[0.0963,0.4569]$ & $0.2020$ & $1.5707$ & $0.2252$ & $0.6580$ & $\HeadW$\\
\bottomrule
\end{tabular}
\caption{The functions $g(x)=Ae^{k\,s(x)}+B$ of the mixture family and the tradeoff functions $c(x)=\nu g(x)$ of Corollary~\ref{thm:numbers}.}\label{tab:bands}
\end{table}

\begin{figure}[htbp]
\centering
\includegraphics[width=\textwidth]{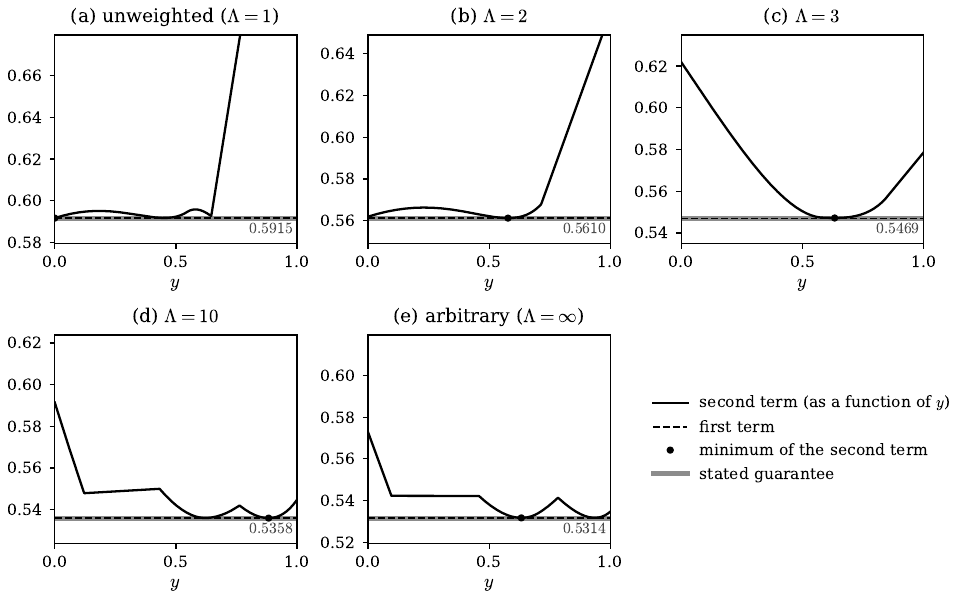}
\caption{The two terms of the guarantee for each row of Table~\ref{tab:bands}: $\alpha(g)$ in \eqref{eq:alpha} for (a), and $\alpha_\Lambda(g,\nu g)$ in \eqref{eq:alphawL} for (b)--(e), with $\Lambda=\infty$ in (e). The solid curve is $\alpha_2(g,y)$, respectively $\alpha_{\Lambda,2}(g,\nu g,y)$, as a function of $y$, and the dashed line is $\alpha_1(g)$, respectively $\alpha_{\Lambda,1}(g,\nu g)$. The guarantee is the smaller of the dashed line and the minimum of the curve. The grey line is at the guarantee stated in Table~\ref{tab:bands}, whose value is printed below it. To help with verification, each panel has been reproduced in Desmos, at \texttt{desmos.com/calculator/} followed by (a)~\href{https://www.desmos.com/calculator/eap9jdvncv}{\texttt{eap9jdvncv}}, (b)~\href{https://www.desmos.com/calculator/s1gsothj9g}{\texttt{s1gsothj9g}}, (c)~\href{https://www.desmos.com/calculator/xbtx3cjgu0}{\texttt{xbtx3cjgu0}}, (d)~\href{https://www.desmos.com/calculator/5x3nyprxud}{\texttt{5x3nyprxud}}, (e)~\href{https://www.desmos.com/calculator/ozz8p9tthx}{\texttt{ozz8p9tthx}}.}\label{fig:bands}
\end{figure}

\begin{corollary}\label{thm:numbers}
Consider the function $g$, the constant $\nu$ and the parameter $\Lambda$ in a row of Table~\ref{tab:bands}. On instances whose weights lie within the factor $\Lambda$, the competitive ratio of Perturbed Greedy with perturbation function $g$ is lower bounded by the last entry of the row. In particular:
\begin{enumerate}
\item In the unweighted setting, \PG\ with band $[\lo,\hi]=[0.6457,1]$, $A=0.0470$, $k=3.1813$ and $B=0.3613$ is at least $\HeadU$-competitive.
\item In the vertex-weighted setting, \PG\ with band $[\lo,\hi]=[0.0963,0.4569]$, $A=0.2020$, $k=1.5707$ and $B=0.2252$ is at least $\HeadW$-competitive (Theorem~\ref{thm:weighted} with $\Lambda=\infty$ and $c(x)=0.6580\,g(x)$).
\item With weights within a factor $\Lambda=2$, $3$ or $10$ of each other, \PG\ with the band of the corresponding row is at least $0.5610$-, $0.5469$- or $0.5358$-competitive, respectively (Theorem~\ref{thm:weighted} with $c(x)=\nu g(x)$).
\end{enumerate}
\end{corollary}
\begin{proof}
By Theorem~\ref{thm:main} for the first row, and by Theorem~\ref{thm:weighted} with $c(x)=\nu g(x)$ for the other rows, it suffices to show that $\alpha(g)$, respectively $\alpha_\Lambda(g,c)$, is at least the stated value. We give a ``proof by picture'' for each row in Figure~\ref{fig:bands}. Recall that $\alpha(g)$ is the smaller of $\alpha_1(g)$ and the minimum over $y$ of $\alpha_2(g,y)$, and similarly for $\alpha_\Lambda(g,c)$. For the first row, the figure plots $\alpha_1(g)$ (dashed) and $\alpha_2(g,y)$ as a function of $y$ (solid), and for the other rows it plots $\alpha_{\Lambda,1}(g,c)$ (dashed) and $\alpha_{\Lambda,2}(g,c,y)$ (solid). The guarantee is the smaller of the dashed line and the minimum of the solid curve, and in every panel it is at least the stated value (grey line). 

Formally, each picture reduces to a finite check. For $x\le y$, the integrand of $\alpha_2(g,y)$ is nonnegative (property~(P3) in the proof of Lemma~\ref{lem:integrate}) and nondecreasing in $g(y)$, so the integral in $\alpha_2(g,y)$ is nondecreasing in $y$, and $\alpha_2(g,y)\ge\alpha_2(g,y')-(g(y'')-g(y'))$ for $y'\le y\le y''$; the same holds for $\alpha_{\Lambda,2}(g,c,y)$. We evaluated each curve on a uniform grid of $16{,}000$ values of $y$ (together with $\lo$ and $\hi$), and in every row the resulting lower bound on its minimum, as well as the dashed line, exceeds the stated value. To help with verification, each panel in the figure has been reproduced in an online plotting tool (Desmos) and the hyperlinks are included in the figure's caption.
\end{proof}
\section{Upper Bounds from Prior Work}\label{app:ub}
With deterministic edges, \PG\ with $g(x)=e^{x-1}$ is $(1-1/e)$-competitive for vertex-weighted OBM \citep{AGKM11}, and no online algorithm does better \citep{KVV90}. We show that for vertex-weighted OMSR, a combination of prior results rules out the possibility of achieving $1-1/e$ for every perturbation function (Proposition~\ref{prop:noopt}). In other words, every Perturbed Greedy algorithm has competitive ratio strictly less than $1-1/e$. The proof combines a result of \cite{LTXZZ23} for deterministic vertex-weighted instances and Adwords instances with an observation of \cite{Udw24}. 

\paragraph{\PG\ on Deterministic Instances.} \PG\ draws a seed $u_k\sim U[0,1]$ independently for every resource $k$. In vertex-weighted OBM, where all edge probabilities are $1$, it matches each arrival to an available neighbor $k$ with the largest perturbed weight $r_k(1-g(u_k))$. In Adwords, resource $k$ has a budget and a bid $b_{kt}$ for each arrival $t$, and \PG\ allocates $t$ to an available neighbor with the largest perturbed bid $b_{kt}(1-g(u_k))$. A resource becomes unavailable when it runs out of budget. \cite{LTXZZ23} call this algorithm Perturbed Ranking, with the perturbation function $f=1-g$ (their Definition~2.1).

\paragraph{The Result of \cite{LTXZZ23}.} \cite{LTXZZ23} show that there is no perturbation function for which \PG\ is $(1-1/e)$-competitive for Adwords, even in the fractional allocation setting. Their proof uses two families of instances, and each family gives an inequality that $g$ and the competitive ratio $\Gamma$ for this $g$ must satisfy. The first family consists of instances of vertex-weighted OBM, a special case of vertex-weighted OMSR. The second consists of instances of Adwords. We show that there are instances of OMSR that are (almost) equivalent to these hard Adwords instances. Using this, it is not hard to show that the upper bound of \cite{LTXZZ23} also applies to vertex-weighted OMSR.

\begin{lemma}[{\cite[Theorems~1 and~2]{LTXZZ23}}]\label{lem:ltxzz}
	Let $g:[0,1]\to[0,1]$ be nondecreasing and right-continuous, and let $\Gamma>0$. Suppose that \PG\ with $g$ is $\Gamma$-competitive for vertex-weighted OBM and on the hard Adwords instances described below. 
	Then $\Gamma<1-1/e-0.0003$.
\end{lemma}

The hard Adwords instances of \cite{LTXZZ23} have $n+1$ offline resources and $2n$ arrivals. Resources in $[n]$ have budget $1$ each and bid $1$ on all $2n$ arrivals. The $(n+1)$-th resource has unlimited budget and bids only on the first $n$ arrivals. We call it the \emph{heavy-budget resource}. Let $b_t$ denote its bid on arrival $t\in[n]$. The values $\{b_t\}_{t\in[n]}$ are determined by $g$ and two parameters $0\le\beta\le\alpha<1$. Ties between the perturbed bid of the heavy-budget resource and that of any other resource are always resolved against the heavy-budget resource. Observe that apart from the heavy-budget resource, the rest of the instance is simply an instance of OBM. 

 \paragraph{The Result of \cite{Udw24}.} To show an upper bound on the competitive ratio of \PG\ with the classic function $g(x)=e^{x-1}$, \cite{Udw24} showed that the hard instances of Adwords with unknown budgets proposed by \cite{Udw25} can be converted to instances of vertex-weighted OMSR. We simply observe that this transformation also applies to the hard Adwords instances above. Let $I$ be one such instance with some $g$ and values $\alpha, \beta$ (as described above). 
 The key is to convert the heavy-budget resource. For $\varepsilon\in(0,1]$, let $I_\varepsilon$ be the vertex-weighted OMSR instance in which the heavy-budget resource is replaced by a \emph{heavy-weight resource}, of weight $B/\varepsilon$ with an edge of probability $\varepsilon b_t/B$ to each of the first $n$ arrivals $t\in[n]$, where $B>0$ is an upper bound on the bids $b_t$. All other resources in $I$ are taken as is, giving $n$ resources of weight $1$ with edges of probability $1$ to all $2n$ arrivals. Observe that the reduced price of the heavy-weight resource at $t$ with seed $u$ is $(B/\varepsilon)(\varepsilon b_t/B)(1-g(u))=b_t(1-g(u))$, matching the perturbed bid of the heavy-budget resource in $I$. Similarly, the reduced prices of all other resources match their perturbed bids for identical seed values. Further, as $\varepsilon\to0$, a probe of the heavy-weight resource in $I_\varepsilon$ rarely succeeds, so this resource behaves like a resource of unlimited budget. The next lemma makes this precise. It is essentially Lemma~3.5 of \cite{Udw24}, stated there for one such instance and for $g(x)=e^{x-1}$. We simply observe that the proof holds for every $g$ and include it here for completeness.

\begin{lemma}[A Heavy Resource; {\cite[Lemma~3.5]{Udw24}}]\label{lem:heavy}
Let $\PG(I)$ be the value of \PG\ with $g$ on the Adwords instance $I$, breaking ties in the same fixed order (so against the heavy-budget resource), and let $\mathrm{OPT}(I)$ be the offline optimum of $I$. Then
\[
\big|\E[\PG(I_\varepsilon)]-\E[\PG(I)]\big|\ \le\ 2n^2\varepsilon\max\{1,B\}\qquad\text{and}\qquad\E[\opt(I_\varepsilon)]\ \ge\ (1-n\varepsilon)\,\mathrm{OPT}(I).
\]
\end{lemma}
\begin{proof}
Run \PG\ on $I_\varepsilon$ and on $I$ with the same seeds, and measure the value of \PG\ by summing up the expected rewards $r_kp_{kt}$ of its probes (regardless of outcome). Reduced prices equal perturbed bids, and a probe of a resource of weight $1$ always succeeds, just as an allocation to a resource of budget $1$ exhausts it. So the two runs make the same choices and collect the same rewards unless some probe of the heavy-weight resource succeeds. This bad event has probability at most $\sum_{t\in[n]}\varepsilon b_t/B\le n\varepsilon$, and on it both values lie in $[0,2n\max\{1,B\}]$, because there are $2n$ arrivals and every reward is at most $\max\{1,B\}$. This gives the first claim. To see the second claim, observe that the offline benchmark on $I_\varepsilon$ can follow an optimal allocation of $I$, which we may take to be integral because the constraints of $I$ form a bipartite transportation problem (the resources in $[n]$ have unit bids and unit budgets, and the heavy-budget resource has no budget constraint), as long as the heavy-weight resource is unmatched. 
\end{proof}
\begin{restatedproposition}{prop:noopt}
Let $g:[0,1]\to[0,1]$ be nondecreasing. With vertex weights, \PG\ with perturbation function $g$ has competitive ratio at most $(1-1/e-0.0003)$ against the adaptive offline benchmark.
\end{restatedproposition}
\begin{proof}[Proof of Proposition~\ref{prop:noopt}]
Suppose \PG\ with $g$ is $\Gamma$-competitive for vertex-weighted OMSR. In particular, it is $\Gamma$-competitive for vertex-weighted OBM. To apply Lemma~\ref{lem:ltxzz}, we also need (i) that $g$ is right-continuous, and (ii) that \PG\ with $g$ is $\Gamma$-competitive on the hard Adwords instances. For (i), because seeds are uniform, changing $g$ on a set of measure zero does not change the law of the run of \PG\ on any instance. Since $g$ is nondecreasing, it has at most countably many discontinuities, so we may assume that $g$ is right-continuous. For (ii), let $I$ be a hard Adwords instance (as defined previously). By Lemma~\ref{lem:heavy} and the $\Gamma$-competitiveness of \PG\ on $I_\varepsilon$, we have $\E[\PG(I)]\ge\Gamma(1-n\varepsilon)\,\mathrm{OPT}(I)-2n^2\varepsilon\max\{1,B\}$ for every $\varepsilon\in(0,1]$, and letting $\varepsilon\to0$ gives (ii). Lemma~\ref{lem:ltxzz} now gives $\Gamma<1-1/e-0.0003$.
\end{proof}

\section{Non-Adaptive Algorithms}\label{app:nonadaptive}
An online algorithm is \emph{non-adaptive} if it is oblivious to the outcomes of probes. Specifically, in a non-adaptive algorithm the selection at each arrival depends only on the probabilities revealed so far and on the algorithm's internal randomness but not on the outcomes of its earlier probes. Since it does not know which resources have been matched, it may choose a resource that is no longer available. Probing such a resource generates no revenue.

 \cite[Theorem~5]{MP12} show that no non-adaptive algorithm is better than $0.5$-competitive against the expectation LP benchmark even in the unweighted setting. We show that their instance gives the same bound against the weaker clairvoyant benchmark that knows success probabilities of all edges prior to the first probe but is restricted to making probes in the arrival order. Consequently, the same bound holds also against the adaptive offline benchmark that can probe in any (adaptive) order (and therefore its objective value is at least that of the clairvoyant). 

\paragraph{The Instance.} Let $n\ge1$, and let $p\in(0,1/2]$ be such that $1/p$ is an integer. There are $n$ resources and $n/p$ arrivals. The arrivals are partitioned into groups $V_1,\dots,V_n$ of $1/p$ consecutive arrivals each. The groups arrive in numerical order, and the arrivals within a group in arbitrary order. For a bijection $\sigma:[n]\to U$, let $I_\sigma$ be the instance in which every arrival of $V_k$ has an edge to each of the resources $\sigma(k),\sigma(k+1),\dots,\sigma(n)$. So for $k\in[n]$, resource $\sigma(k)$ can be matched only by the arrivals of $V_1,\dots,V_k$. All weights are $1$ and all edges have probability $p$. The expectation LP benchmark has value $n$ for every instance in the family because it can match all arrivals of $V_k$ to $\sigma(k)$, using up exactly the unit capacity of every resource. \cite{MP12} showed that, for every non-adaptive online algorithm, some instance $I_\sigma$ gives it expected value at most $0.5n+o(n)$ as $n\to\infty$ and $p\to0$. 
\begin{proposition}\label{prop:nonadaptive}
For every $\epsilon>0$, no non-adaptive algorithm is $(0.5+\epsilon)$-competitive against the clairvoyant benchmark (and hence against the adaptive offline benchmark).
\end{proposition}

It suffices to show that on the family of instances $I_\sigma$ the clairvoyant benchmark has expected value $n-o(n)$. The idea is to group consecutive groups of arrivals into large chunks. By concentration, a chunk of $m$ groups produces almost $m$ successful probes that can be matched to distinct resources.

\begin{lemma}\label{lem:nonad-opt}
Let $n=m^2$ for an integer $m$. On every $I_\sigma$, there exists a non-anticipative policy that probes the arrivals in their order of arrival and matches at least $n-2n^{3/4}$ resources in expectation.
\end{lemma}
\begin{proof}
Rename the resources so that $\sigma(j)=j$, so that resource $j$ can be matched only by the arrivals of $V_1,\dots,V_j$. Partition the groups into $m$ chunks of $m$ consecutive groups: chunk $r\in[m]$ consists of $V_{(r-1)m+1},\dots,V_{rm}$, and it has $m/p$ arrivals. Partition the resources in the same way into blocks $B_r:=\{(r-1)m+1,\dots,rm\}$ of $m$ resources. The policy probes the arrivals in their order of arrival. It matches the first chunk only to the resources $m+1,\dots,2m$ of $B_2$, which no arrival after $V_{2m}$ can reach, the second chunk only to the resources of $B_3$, and so on. Specifically, for $r=1,\dots,m-1$, it probes each arrival of chunk $r$ with an arbitrary unmatched resource of $B_{r+1}$, as long as there is one. This is possible because every arrival of chunk $r$ belongs to a group $V_k$ with $k\le rm$, so it has an edge to every resource of the next block $B_{r+1}$. Note that the last arrival chunk and the first resource block are ignored by the policy. 

Draw an independent $\mathrm{Bernoulli}(p)$ outcome for every arrival, and let the probe of that arrival succeed if and only if its outcome is $1$. Let $S_r$ be the number of arrivals of chunk $r$ with outcome $1$. Every such arrival matches a new resource of $B_{r+1}$ until all of them are matched, so the policy matches $\min\{S_r,m\}$ resources of $B_{r+1}$. The number $S_r$ is binomial with $m/p$ trials and success probability $p$, so it has mean $m$ and variance $m(1-p)\le m$. Hence using Cauchy-Schwarz,
\[
\E\big[\min\{S_r,m\}\big]\ =\ m-\E\big[(m-S_r)^+\big]\ \ge\ m-\big(\E\big[(S_r-m)^2\big]\big)^{1/2}\ \ge\ m-\sqrt m .
\]
Summing over $r=1,\dots,m-1$, the policy matches at least $(m-1)(m-\sqrt m)\ge m^2-m-m^{3/2}\ge n-2n^{3/4}$ resources in expectation.
\end{proof}

\section{Away from Ties, Greed is Good}\label{app:greedy}
We formalize the observation in Section~\ref{sec:overview} that Greedy is better than $0.5$-competitive on instances where we are far away from ties. Recall that Greedy probes each arrival $t$ with the available resource of the largest expected reward $r_ip_{it}$. 

\paragraph{Instances That Are Away from Ties.} For $\eta\ge0$, we say that an instance is \emph{$\eta$-separated} if at every arrival the expected rewards of any two edges differ by a factor of at least $1+\eta$, i.e., $r_kp_{kt}\ge(1+\eta)\,r_ip_{it}$ for any two resources $k\ne i$ with an edge to $t$ and $r_kp_{kt}\ge r_ip_{it}$. Arbitrary instances are $0$-separated, and we say that an instance is away from ties if $\eta>0$. 

\begin{proposition}\label{prop:greedy}
On $\eta$-separated instances, Greedy is $\frac{1+\eta}{2+\eta}$-competitive against the adaptive offline benchmark.
\end{proposition}
For $\eta=0$ this is the classical guarantee of $0.5$, and for $\eta\in[0,1]$ the guarantee is $0.5+\frac{\eta}{2(2+\eta)}\ge0.5+\eta/6$. As $\eta\to\infty$, the guarantee tends to $1$, so Greedy is asymptotically optimal against the adaptive offline benchmark. This is not true against the expectation LP benchmark: consider a single resource with an edge of probability $p$ to each of $1/p$ arrivals. Every arrival has a single edge, so the instance is $\eta$-separated for every $\eta$. However, the expectation LP has value $1$, whereas every policy matches the resource with probability at most $1-(1-p)^{1/p}$, which tends to $1-1/e$ as $p\to0$. 

We will prove the proposition using the LP-free approach of \cite{GIU} that we also use to analyze \PG. The proof is meant to be self-contained, and we start by (re)introducing the setup and essential notation below.

\paragraph{Shared Outcomes.} We work in the standard independent model of OMSR where edge $(i,t)$ succeeds with probability $p_{it}$ independent of all else. Let $\opt$ be an optimal offline policy, which we may take to be a deterministic dynamic program without loss of generality. We couple all the edge realizations in $\opt$ and Greedy. 

Let $O_i$ be the set of arrivals at which $\opt$ probes $i$. Since $\opt$ is unaware of the outcome of an edge before probing it, whether some arrival $t$ is in  $O_i$ is determined only by the outcomes of the edges of the other arrivals. Greedy's available resources at $t$, and hence its probe at $t$, are determined fully by the outcomes of the edges probed before arrival $t$.  

\paragraph{The LP-Free Certificate.} Instead of writing an LP relaxation of $\opt$, we split the reward of Greedy on every sample path into nonnegative amounts, $\lambda_t$ for every arrival $t$ and $\theta_k$ for every resource $k$, so that $\sum_t\lambda_t+\sum_k\theta_k$ is the reward of Greedy. Then we compare these amounts with $\opt$ one resource at a time. If for some $\alpha$ and every resource $i$,
\begin{equation}\label{eq:greedyres}
\E\Big[\sum_{t\in O_i}\lambda_t+\theta_i\Big]\ \ge\ \alpha\,r_i\Pr[\opt\text{ matches }i],
\end{equation}
then Greedy is $\alpha$-competitive. Indeed, every arrival lies in at most one set $O_i$ and $\lambda\ge0$, so summing \eqref{eq:greedyres} over $i$ gives that the expected reward of Greedy, $\E[\sum_t\lambda_t+\sum_i\theta_i]$, is at least $\alpha\sum_ir_i\Pr[\opt\text{ matches }i]$, the expected reward of $\opt$ times $\alpha$.

\paragraph{The Dual Solution.} We split the reward of every match of Greedy between the arrival and the resource, giving more than half of the reward to the resource when $\eta>0$. Let $\alpha:=\frac{1+\eta}{2+\eta}$, and notice that $(1-\alpha)(1+\eta)=\alpha$. Initialise all duals to $0$, and if Greedy matches $k$ at $t$, set $\lambda_t:=(1-\alpha)\,r_k$ and $\theta_k:=\alpha\,r_k$. Then $\sum_t\lambda_t+\sum_k\theta_k$ equals the reward of Greedy on every sample path.

\begin{proof}[Proof of Proposition~\ref{prop:greedy}]
By the LP-free certificate, it suffices to show \eqref{eq:greedyres} for every resource $i$.

Fix $i$, and for every arrival $t$ with an edge to $i$ let $\phi_t:=\Pr[t\in O_i]$ be the probability that $\opt$ probes $i$ at $t$. Therefore, $\opt$ matches $i$ at $t$ with probability $\phi_tp_{it}$, and since $\opt$ stops probing $i$ after it matches $i$,
\[
\Pr[\opt\text{ matches }i]\ =\ \sum_t\phi_tp_{it}.
\]
For the same reason, we have 
\[\theta_i\ge\sum_t\one\{t\in O_i,\ (i,t)\text{ succeeds}\}\,\theta_i.\] 
Hence the left side of \eqref{eq:greedyres} is at least $\sum_t\E\big[\one\{t\in O_i\}\big(\lambda_t+\one\{(i,t)\text{ succeeds}\}\,\theta_i\big)\big]$, and it suffices to show that for every arrival $t$,
\begin{equation}\label{eq:greedyarr}
\E\Big[\one\{t\in O_i\}\big(\lambda_t+\one\{(i,t)\text{ succeeds}\}\,\theta_i\big)\Big]\ \ge\ \alpha\,r_i\,\phi_t\,p_{it}.
\end{equation}

To show \eqref{eq:greedyarr}, fix the outcomes of the edges of all arrivals other than $t$. They determine whether $t\in O_i$, whether Greedy has matched $i$ before $t$, and Greedy's probe at $t$, while the edges of $t$ still succeed independently with their probabilities. Let $t\in O_i$. There are three cases.
\begin{itemize}[leftmargin=*]
\item \emph{Greedy has matched $i$ before $t$.} Then $\theta_i=\alpha r_i$, and the expectation of $\lambda_t+\one\{(i,t)\text{ succeeds}\}\,\theta_i$ is at least $\alpha\,r_ip_{it}$.
\item \emph{$i$ is available at $t$ and Greedy probes $i$ at $t$.} Greedy probes the same edge as $\opt$ and observes the same outcome. So if $(i,t)$ succeeds, Greedy matches $i$ at $t$, and $\lambda_t+\theta_i=r_i$. The expectation is at least $r_ip_{it}$.
\item \emph{$i$ is available at $t$ and Greedy probes a resource $k\ne i$ at $t$.} Then $r_kp_{kt}\ge r_ip_{it}$ because Greedy prefers $k$ to $i$, and hence $r_kp_{kt}\ge(1+\eta)\,r_ip_{it}$ because the instance is $\eta$-separated. Greedy matches $k$ at $t$ if $(k,t)$ succeeds, so the expectation of $\lambda_t$ is $(1-\alpha)\,r_kp_{kt}\ge(1-\alpha)(1+\eta)\,r_ip_{it}=\alpha\,r_ip_{it}$.
\end{itemize}
In every case, the expectation over the outcomes of the edges of $t$ is at least $\alpha\,r_ip_{it}$. Averaging over the other outcomes gives \eqref{eq:greedyarr}, since $t\in O_i$ with probability $\phi_t$. Summing \eqref{eq:greedyarr} over $t$ gives \eqref{eq:greedyres}.
\end{proof}

\section{Deferred Proofs}\label{app:deferred}

\paragraph{Lemma~\ref{lem:integrate}.}\label{app:integrate} \emph{For every $\pi\ge0$, $\int_0^1\gamma(x,\pi)\,dx\ge\alpha(g)$ and $\int_0^1\gamma_\Lambda(x,\pi)\,dx\ge\alpha_\Lambda(g,c)$. Both bounds are tight: $\alpha(g)$ and $\alpha_\Lambda(g,c)$ are the infima of the two integrals over $\pi\ge0$.}
\begin{proof}
We focus on the claims for $\gamma$. As we discuss later, essentially the same proof applies to $\gamma_\Lambda$. Recall that $g$ is continuous and nondecreasing with $g(1)<1$. By the definition of the target $\gamma$ in Section~\ref{sec:dual},
\[
\gamma(x,\pi)=\pi+\one\{1-g(x)>\pi\}\,h(x,\pi),\qquad\text{where}\qquad h(x,\pi):=\min\{g(x),\ \sigma(x)g(x)+1-g(x)-\pi\}.
\]
Recall that $\alpha(g)=\min\{\alpha_1(g),\ \min_{y\in[0,1]}\alpha_2(g,y)\}$, where
\begin{align*}
\alpha_1(g)&=\int_0^1\min\{g(x),\ \sigma(x)g(x)+1-g(x)\}\,dx,\\
\alpha_2(g,y)&=1-g(y)+\int_0^y\min\{g(x),\ \sigma(x)g(x)+g(y)-g(x)\}\,dx.
\end{align*}
Rewriting $\alpha_1$ and $\alpha_2$ in terms of $h$, we have
\begin{equation}\label{eq:alphah}
\alpha_1(g)=\int_0^1h(x,0)\,dx\qquad\text{and}\qquad\alpha_2(g,y)=1-g(y)+\int_0^yh\big(x,1-g(y)\big)\,dx .
\end{equation}
Let $I(\pi):=\int_0^1\gamma(x,\pi)\,dx$. We show that $\inf_{\pi\ge0}I(\pi)=\alpha(g)$, which gives both claims.

\emph{Extending $g$ and $\alpha_2$.} To treat the two terms of $\alpha(g)$ alike, we extend $g$ to $[0,2]$ by setting $g(2):=1$ and interpolating linearly between $g(1)$ and $g(2)$. The extended function $g$ is continuous and nondecreasing on $[0,2]$. Similarly, we extend the domain of $\alpha_2(g,\cdot)$ to $[0,2]$ by letting
\[
\alpha_2(g,y):=1-g(y)+\int_0^{\min\{1,y\}}h\big(x,1-g(y)\big)\,dx\qquad\text{for }y\in[0,2],
\]
which agrees with \eqref{eq:alphah} for $y\in[0,1]$. Since $g(2)=1$, this extension captures the first term: $\alpha_2(g,2)=\int_0^1h(x,0)\,dx=\alpha_1(g)$.

We first show that the minimum of the extended function $\alpha_2(g,\cdot)$ is the guarantee $\alpha(g)$. Then we show that the infimum of $I$, the integral of the target, matches this minimum.

\emph{Step 1: We show that $\min_{y\in[0,2]}\alpha_2(g,y)=\alpha(g)$.} For $y\in[1,2]$, $\alpha_2(g,y)=\pi+\int_0^1h(x,\pi)\,dx$ with $\pi=1-g(y)$. Since $h(x,\pi)$ is the minimum of two affine functions of $\pi$, this is a concave function of $g(y)\in[g(1),1]$. So its minimum over $y\in[1,2]$ is attained at $y=1$ or $y=2$, where it equals $\alpha_2(g,1)$ and $\alpha_2(g,2)=\alpha_1(g)$, respectively. Hence
\[
\min_{y\in[0,2]}\alpha_2(g,y)=\min\Big\{\alpha_1(g),\ \min_{y\in[0,1]}\alpha_2(g,y)\Big\}=\alpha(g).
\]

\emph{Step 2: We show that $\inf_{\pi\ge0}I(\pi)=\min_{y\in[0,2]}\alpha_2(g,y)$.} For $\pi>1-g(0)$ the indicator vanishes for every $x$, so $I(\pi)=\pi>I(1-g(0))$. Therefore, the infimum of $I$ over $\pi\ge0$ equals its infimum over $[0,1-g(0)]$. By the intermediate value theorem, every $\pi\in[0,1-g(0)]$ can be expressed as $\pi=1-g(y)$ for some $y\in[0,2]$, because the extended $g$ is continuous with $g(0)\le1-\pi\le1=g(2)$. Now fix $y\in[0,2]$ and let $\pi=1-g(y)$. Then the indicator $\one\{1-g(x)>\pi\}=\one\{g(x)<g(y)\}$. Since $g$ is nondecreasing, the indicator vanishes for $x\ge y$ and we have
\[
I\big(1-g(y)\big)=1-g(y)+\int_0^{\min\{1,y\}}\one\{g(x)<g(y)\}\,h\big(x,1-g(y)\big)\,dx\ \le\ \alpha_2(g,y).
\]
Here the inequality follows from the non-negativity of $h(x,1-g(y))=\min\{g(x),\ \sigma(x)g(x)+g(y)-g(x)\}\ge0$ for $x\le y$. If $y$ is the smallest point at which $g$ takes the value $g(y)$, then $g(x)<g(y)$ for all $x<y$, and the inequality turns into an equality, i.e., $I(1-g(y))=\alpha_2(g,y)$. So every value of $I$ on $[0,1-g(0)]$ matches a value of $\alpha_2(g,\cdot)$ (at the smallest $y$ with $g(y)=1-\pi$, which exists because $g$ is continuous), and every value of $\alpha_2(g,\cdot)$ is at least a value of $I$. Hence the infima of both functions coincide.

Combining Steps~1 and~2, $\inf_{\pi\ge0}I(\pi)=\alpha(g)$.

\emph{The weighted target.} For $\gamma_\Lambda$, we simply replace $h$ by
\[
h_\Lambda(x,\pi):=\min\{g(x)-c(x),\ 1-g(x)-\pi+(g(x)-c(x))\,\sigma_\Lambda(x)\}+c(x)\,w(x,\pi),
\]
so that $\gamma_\Lambda(x,\pi)=\pi+\one\{1-g(x)>\pi\}\,h_\Lambda(x,\pi)$. The proof above used only three properties of $h$:
\begin{enumerate}[label=(P\arabic*)]
\item $\alpha_1(g)$ and $\alpha_2(g,y)$ can be written in terms of $h$ as in \eqref{eq:alphah};
\item $h(x,\pi)$ is concave in $\pi$ (Step~1);
\item $h(x,1-g(y))\ge0$ for $x\le y$ (Step~2).
\end{enumerate}
The function $h_\Lambda$ has the same properties, with $\alpha_{\Lambda,1}(g,c)$ and $\alpha_{\Lambda,2}(g,c,y)$ in place of $\alpha_1(g)$ and $\alpha_2(g,y)$. For (P1), $w(x,0)=1$ because $\sigma_\Lambda(x)\le\sigma(x)\le1$, so substituting $\pi=0$ and $\pi=1-g(y)$ in $h_\Lambda$ gives exactly the integrands of $\alpha_{\Lambda,1}(g,c)$ and $\alpha_{\Lambda,2}(g,c,y)$ in \eqref{eq:alphawL}. For (P2) and (P3), observe that $h_\Lambda(x,\pi)$ is the sum of two minima of affine functions of $\pi$, the second multiplied by $c(x)\ge0$, so it is concave in $\pi$. At $\pi=1-g(y)$ with $x\le y$, each of these affine functions is nonnegative because $0\le c(x)\le g(x)\le g(y)$, so $h_\Lambda(x,1-g(y))\ge0$.
Therefore Steps~1 and~2 apply verbatim, with $\alpha_{\Lambda,2}(g,c,\cdot)$ extended to $[0,2]$ in the same way, and give $\inf_{\pi\ge0}\int_0^1\gamma_\Lambda(x,\pi)\,dx=\alpha_\Lambda(g,c)$.
\end{proof}

\paragraph{Lemma~\ref{lem:transform}.} \emph{For every non-anticipative policy, the joint law of its sequence of probes and outcomes, and hence the law of its value, is the same in the independent and the correlated models.}
\begin{proof}
Let $\xi$ denote the internal randomness of the policy. Conditioned on $\xi$, we show by induction on $k$ that the law of the first $k$ probes and outcomes is the same in the two models. This is trivially true for $k=0$. Fix a history $h=((e_1,o_1),\dots,(e_k,o_k))$ observed by the policy, where $e_\ell=(i_\ell,t_\ell)$ is the $\ell$-th edge probed and $o_\ell$ its outcome. The arrivals $t_\ell$ are distinct, because each arrival is probed at most once. In the correlated model, the event that the history equals $h$ is determined by $\xi$ and $(z_{t_1},\dots,z_{t_k})$ only. Suppose the next probe is $e=(i,t)$; it is a function of $(\xi,h)$, and $t$ is distinct from $t_1,\dots,t_k$ for the same reason. Since $z_t$ is independent of $(z_{t_1},\dots,z_{t_k})$, we get $\Pr[z_t\le p_{it}\mid h]=p_{it}$, which is the success probability in the independent model. Therefore the two models assign the same probability to every history of length $k+1$.
\end{proof}
\paragraph{Lemma~\ref{lem:cont}.} \emph{Pathwise, at every arrival $t$, the set of resources available in $\alg$ at $t$ contains the set available in $\algmi$ at $t$. Consequently $P_t\ge P^{-i}_t$ and $e_t\geq 0$ for all $t$.}
\begin{proof}
We prove the claim by induction over arrivals. It holds at the first arrival, where the two sets are $U$ and $U\setminus\{i\}$. Let $A\supseteq B$ be the sets available in $\alg$ and in $\algmi$ at an arrival $t$, and let $\alg$ probe $j$ and $\algmi$ probe $k$ (either may probe nothing). If $j\in B$, deterministic tie breaking gives $k=j$, both runs see the same outcome $\one\{z_t\le p_{jt}\}$, and containment is preserved. If $j\notin B$ or $\alg$ probes nothing, then $A$ and $B$ each lose at most one resource: $A$ loses at most $j\in A\setminus B$, and $B$ loses at most $k$. Since every element of $B$ lies in $A\setminus\{j\}$, we have $B\setminus\{k\}\subseteq A\setminus\{j\}$, and containment is preserved. The claim on prices follows because $P_t$ is a maximum over a superset of the set over which $P^{-i}_t$ is a maximum, and then $e_t=(P_t-P^{-i}_t)/p_{it}\ge0$.
\end{proof}

\paragraph{Lemma~\ref{lem:floor}.} \emph{$\E\big[\sum_{t\in O_i}\lambda_t\big]=\E\big[\sum_{t\in O_i}\one\{z_t\le p_{it}\}(\pi_t+e_t)\big]$.}
\begin{proof}
Recall that the seeds are fixed. Consider an arbitrary arrival $t$ and thresholds $\bz_{-t}$. It suffices to show that
\begin{equation}\label{eq:floorchain}
\E\big[\one\{t\in O_i\}\lambda_t\,\big|\,\bz_{-t}\big]\ =\ \one\{t\in O_i\}\,p_{it}(\pi_t+e_t)\ =\ \E\big[\one\{t\in O_i\}\one\{z_t\le p_{it}\}(\pi_t+e_t)\,\big|\,\bz_{-t}\big],
\end{equation}
and then to take expectations and sum over $t$. Observe that the set of resources available in $\alg$ at $t$, the prices $P_t$ and $P^{-i}_t$, and hence $\pi_t$ and $e_t$, are functions of $\bz_{<t}$ alone. Let $k$ denote the resource probed by $\alg$ at $t$ (if there is none, then $\lambda_t=0=P_t$). By definition $\lambda_t=(1-g(u_k))\one\{z_t\le p_{kt}\}$, and $z_t$ is uniform on $[0,1]$ and independent of $\bz_{-t}$, so
\[
\E[\lambda_t\mid\bz_{-t}]=(1-g(u_k))\,p_{kt}=P_t=p_{it}(\pi_t+e_t).
\]
Since $\opt$ is deterministic and non-anticipative, $\one\{t\in O_i\}$ is determined by $\bz_{-t}$: if $\bz'$ agrees with $\bz$ except in coordinate $t$, the runs of $\opt$ on $\bz$ and $\bz'$ coincide up to and including the edge of $t$ that it probes, because until then no outcome that depends on $z_t$ has been observed. This gives the first equality of \eqref{eq:floorchain}. The second follows because $\E[\one\{z_t\le p_{it}\}\mid\bz_{-t}]=p_{it}$ and $\pi_t+e_t$ is determined by $\bz_{<t}$.
\end{proof}

\section{The Unweighted Argument with Weights}\label{app:Ip}

This appendix rederives the statements of Section~\ref{sec:analysis} with vertex weights. We first update the notation and then generalize Sections~\ref{sec:struct} and~\ref{sec:relarr}. 

\paragraph{Notation.} Fix $i$, $u=u_i$ and the seeds $\bu_{-i}$, and let all weights lie within a factor $\Lambda\in[1,\infty]$ of each other, so that $\kappa_j\ge\sigma_\Lambda(u)$ for every $j\ne i$. As in Section~\ref{sec:analysis}, all seeds are fixed, and $\E$ and $\Pr$ are over the thresholds $\bz$. The reduced price of $j$ at $t$ is $r_jp_{jt}(1-g(u_j))$, and $P_t$ and $P^{-i}_t$ are the reduced prices of the probes of $\alg$ and $\algmi$ at $t$ ($0$ if none). As in Section~\ref{sec:weighted}, we normalize reduced prices by $r_ip_{it}$. For every arrival $t$ with an edge to $i$,
\[
\pi_t:=\frac{P^{-i}_t}{r_ip_{it}},\qquad\delta_t(u):=1-g(u)-\pi_t,\qquad e_t:=\frac{P_t-P^{-i}_t}{r_ip_{it}},
\]
so that $P_t=r_ip_{it}(\pi_t+e_t)$. If $\alg$ probes $k$ at $t$, then $\lambda_t=r_k(1-g(u_k))\one\{z_t\le p_{kt}\}$, and $\theta_i\in\{0,r_ig(u)\}$.

Lemma~\ref{lem:cont} holds verbatim with the same proof which uses only that \PG\ probes an available resource of largest reduced price and has deterministic tie breaking. Thus, $e_t\ge0$. Lemma~\ref{lem:floor} holds with the factor $r_i$ as stated below.

\begin{lemma}[Weighted Form of Lemma~\ref{lem:floor}]\label{lem:wfloor}
$\E\big[\sum_{t\in O_i}\lambda_t\big]=r_i\,\E\big[\sum_{t\in O_i}\one\{z_t\le p_{it}\}(\pi_t+e_t)\big]$.
\end{lemma}
\begin{proof}
Fix an arrival $t$ and thresholds $\bz_{-t}$. As in the proof of Lemma~\ref{lem:floor}, the resource $k$ that $\alg$ probes at $t$, the prices $P_t$ and $P^{-i}_t$, and hence $\pi_t$ and $e_t$, are functions of $\bz_{<t}$. The indicator $\one\{t\in O_i\}$ is a function of $\bz_{-t}$. Since $z_t$ is uniform on $[0,1]$ and independent of $\bz_{-t}$,
\[
\E[\lambda_t\mid\bz_{-t}]=r_k(1-g(u_k))\,p_{kt}=P_t=r_ip_{it}(\pi_t+e_t)
\]
(both sides are $0$ if $\alg$ probes nothing at $t$), and $\E[\one\{z_t\le p_{it}\}\mid\bz_{-t}]=p_{it}$. Hence
\[
\E\big[\one\{t\in O_i\}\lambda_t\,\big|\,\bz_{-t}\big]=\one\{t\in O_i\}\,r_ip_{it}(\pi_t+e_t)=r_i\,\E\big[\one\{t\in O_i\}\one\{z_t\le p_{it}\}(\pi_t+e_t)\,\big|\,\bz_{-t}\big].
\]
Taking expectations and summing over $t$ gives the claim.
\end{proof}
\subsection{What We Need from \texorpdfstring{$\theta_i$}{theta\_i}}\label{app:Ipstruct}
For ease of reference, we recall the definitions of Section~\ref{sec:wstruct}. At an arrival $t$ with $(i,t)\in E$, the demand is
\begin{align*}
D_t&=\big(\gamma_\Lambda(u,\pi_t)-(\pi_t+e_t)\big)^+\\
&=\Big(\one\{\delta_t(u)>0\}\big(\min\{g(u)-c(u),\ (g(u)-c(u))\sigma_\Lambda(u)+\delta_t(u)\}+c(u)\,w_t(u)\big)-e_t\Big)^+.
\end{align*}
Let $v_t(u)=\sigma_\Lambda(u)$ if $\alg$'s probe at $t$ has reduced price at least $r_ip_{it}(1-g(u))$, and $v_t(u)=1$ otherwise. The maximum demand is $M=\max_{t\in T_i}D_t$, where $T_i=\{t:\ (i,t)\in E,\ \delta_t(u)>0,\ z_t\le p_{it}\}$. The counterpart of Lemma~\ref{lem:bounds} is the following.

\begin{restated}{Lemma}{lem:clinear}
Pathwise, for every $t$ with $(i,t)\in E$ and $\delta_t(u)>0$:
\begin{enumerate}[label=(\Alph*)]
\item $D_t\le g(u)-c(u)+c(u)\,w_t(u)$;
\item if $\alg$ probes at $t$ a resource whose reduced price is at least $r_ip_{it}(1-g(u))$, then $e_t\ge\delta_t(u)$ and $D_t\le(g(u)-c(u))\sigma_\Lambda(u)+c(u)\,w_t(u)$.
\end{enumerate}
Overall, $D_t\le(g(u)-c(u))\,v_t(u)+c(u)\,w_t(u)$.
\end{restated}
\begin{proof}[Proof of Lemma~\ref{lem:clinear}]
Let $t$ have $(i,t)\in E$ and $\delta_t(u)>0$, so that
\[
D_t=\big(\min\{g(u)-c(u),\ (g(u)-c(u))\sigma_\Lambda(u)+\delta_t(u)\}+c(u)\,w_t(u)-e_t\big)^+ .
\]
For (A), the minimum is at most $g(u)-c(u)$, and $e_t\ge0$ by Lemma~\ref{lem:cont}, so $D_t\le g(u)-c(u)+c(u)\,w_t(u)$. For (B), the reduced price of $\alg$'s probe at $t$ is $P_t=r_ip_{it}(\pi_t+e_t)$, so $P_t\ge r_ip_{it}(1-g(u))$ gives $e_t\ge1-g(u)-\pi_t=\delta_t(u)$. Bounding the minimum by its second term then gives
\[
D_t\ \le\ \big((g(u)-c(u))\sigma_\Lambda(u)+\delta_t(u)+c(u)\,w_t(u)-\delta_t(u)\big)^+\ =\ (g(u)-c(u))\sigma_\Lambda(u)+c(u)\,w_t(u).
\]
For the last claim, if $\alg$'s probe at $t$ has reduced price at least $r_ip_{it}(1-g(u))$, then $v_t(u)=\sigma_\Lambda(u)$ and (B) applies; otherwise $v_t(u)=1$ and (A) applies.
\end{proof}
The counterpart of Lemma~\ref{lem:red} is the following.

\begin{restated}{Lemma}{lem:wred}
If $\E[M]\le \frac{1}{r_i}\E[\theta_i]=g(u)\Pr[\alg\text{ matches }i]$, then \eqref{eq:perseed} holds for the target $\gamma_\Lambda$.
\end{restated}
\begin{proof}[Proof of Lemma~\ref{lem:wred}]
By Lemma~\ref{lem:wfloor}, the per-seed inequality \eqref{eq:perseed} for the target $\gamma_\Lambda$ is equivalent to
\[
\E[\theta_i]\ \ge\ r_i\,\E\Big[\sum_{t\in O_i}\one\{z_t\le p_{it}\}\big(\gamma_\Lambda(u,\pi_t)-\pi_t-e_t\big)\Big].
\]
There is at most one $t\in O_i$ with $z_t\le p_{it}$, because $\opt$ consumes $i$ at its first success on $i$. Hence, pathwise,
\[
\sum_{t\in O_i}\one\{z_t\le p_{it}\}\big(\gamma_\Lambda(u,\pi_t)-\pi_t-e_t\big)\ \le\ \sum_{t\in O_i}\one\{z_t\le p_{it}\}\,D_t\ \le\ M,
\]
and the hypothesis $\E[\theta_i]\ge r_i\,\E[M]$ gives the inequality above.
\end{proof}

\subsection{An ``Almost'' Pathwise Proof of \texorpdfstring{Lemma~\ref{prop:Ip}(A)}{part (A)}}\label{app:IpA}
The following definition, lemma and corollary of Section~\ref{sec:relarr} hold verbatim, with the same proofs.

\begin{restateddefinition}{def:release}
If $\alg$ probes $i$ at $\tau$, its \emph{best alternative} $J=J_\tau$ is the resource $\alg$ would probe at $\tau$ if $i$ were removed from its current available set (if any). We say $\alg$ \emph{releases} at $\tau$ if it unsuccessfully probes $i$ although a probe of $J$ would have succeeded. Formally, the best alternative $J$ is released at $\tau$ if $\alg$ probes $i$ at $\tau$, $J$ exists, and $p_{i\tau}<z_\tau\le p_{J\tau}$.
\end{restateddefinition}

\begin{restated}{Lemma}{lem:shadow}
Consider two runs of \PG\ from some arrival $s$ onward, with the same seeds and thresholds, starting from available sets $S\cup\{i\}$ (run~1) and $S$ (run~2), where $i\notin S$. Suppose run~1 neither matches $i$ nor releases at any arrival $t\ge s$. Then at every arrival $t\ge s$ the two runs have the same available resources other than $i$. Moreover, at every arrival where run~1 does not probe $i$, both runs probe the same resource with the same outcome.
\end{restated}

\begin{restated}{Corollary}{cor:norelease}
If $\alg$ never matches $i$ and never releases, then $T_i=\emptyset$ and $M=0$.
\end{restated}
 Recall from Section~\ref{sec:relarr} the threshold vector $\bz^{(\tau)}$ and the run of $\alg$ in which $(i,\tau)$ succeeds. The definition of a release arrival generalizes as follows.

\begin{definition}[Release Arrival with Weights]\label{def:wrelarr}
An arrival $\tau$ is a \emph{release arrival} if
\begin{enumerate}[label=(\roman*)]
\item $\alg$ probes $i$ at $\tau$, a best alternative $J=J_\tau$ exists, and $p_{i \tau}<p_{J\tau}$, and
\item in the run of $\alg$ in which $(i,\tau)$ succeeds, at every arrival $t>\tau$ with $(i,t)\in E$ and $z_t\le p_{it}$, $\alg$ probes a resource whose reduced price is at least $r_ip_{it}(1-g(u))$.
\end{enumerate}
\end{definition}

The following lemma holds for this definition with the same proof, which does not involve reduced prices.

\begin{restated}{Lemma}{lem:meas}
Condition~(i) of Definition~\ref{def:wrelarr}, the resource $J_\tau$ and the probability $p_{J\tau}$ are determined by $\bz_{<\tau}$, and condition~(ii) is determined by $\bz_{-\tau}$. In particular, whether $\tau$ is a release arrival does not depend on $z_\tau$.
\end{restated}
Recall from Section~\ref{sec:wstruct} that $V=\max_{t\in T_i}v_t(u)$, with $\max\emptyset:=0$. Since $\sigma_\Lambda(u)\le1$, we have $v_t(u)\le1$. Moreover, $v_t(u)=\sigma_\Lambda(u)$ whenever $i$ is available in $\alg$ at $t$, because $\alg$ then probes a resource whose reduced price is at least that of $i$.

\paragraph{The Auxiliary Dual.} Lemma~\ref{prop:Ip}(A) compares $\E[V]$ with $\Pr[\alg\text{ matches }i]$. Since 
\[\Pr[\alg\text{ matches }i]=\frac{\E[\theta_i]}{r_ig(u)},\] 
we define a normalized auxiliary dual that equals $\Pr[\alg\text{ matches }i]$ in expectation. For every arrival $\tau$, let $a_\tau$ be the indicator that $\alg$ probes $i$ at $\tau$ and $z_\tau\le p_{i\tau}$, so that $\sum_\tau a_\tau=\one\{\alg\text{ matches }i\}$, and let
\[
\hat a_\tau:=\begin{cases}
p_{i\tau}/p_{J\tau} & \text{if $\tau$ is a release arrival and } z_\tau\le p_{J\tau},\\
0 & \text{if $\tau$ is a release arrival and } z_\tau>p_{J\tau},\\
a_\tau & \text{otherwise.}
\end{cases}
\]

\begin{lemma}[Weighted Form of Lemma~\ref{lem:mean}]\label{lem:wmean}
For every arrival $\tau$, $\E[\hat a_\tau\mid\bz_{-\tau}]=\E[a_\tau\mid\bz_{-\tau}]$. Consequently, $\E[\sum_\tau\hat a_\tau]=\Pr[\alg\text{ matches }i]$.
\end{lemma}
\begin{proof}
Fix $\bz_{-\tau}$. By Lemma~\ref{lem:meas}, this determines whether $\tau$ is a release arrival, and also $J_\tau$ and $p_{J\tau}$. The threshold $z_\tau$ is uniform on $[0,1]$ and independent of $\bz_{-\tau}$. If $\tau$ is not a release arrival, then $\hat a_\tau=a_\tau$. Otherwise $\alg$ probes $i$ at $\tau$, so $\E[a_\tau\mid\bz_{-\tau}]=p_{i\tau}$ and $\E[\hat a_\tau\mid\bz_{-\tau}]=(p_{i\tau}/p_{J\tau})\,p_{J\tau}=p_{i\tau}$. The second claim follows by taking expectations and summing over $\tau$.
\end{proof}

\begin{proof}[Proof of Lemma~\ref{prop:Ip}(A)]
By Lemma~\ref{lem:wmean}, it suffices to show $\E[\sum_\tau\hat a_\tau]\ge\E[V]$. We show that $\sum_\tau\hat a_\tau\ge V$ pathwise.

If $\alg$ probes $i$ at $\tau$ with best alternative $J$, then $r_ip_{i\tau}(1-g(u))\ge r_Jp_{J\tau}(1-g(u_J))$. Dividing by $r_ip_{J\tau}(1-g(u))$ gives the weighted \emph{crossing} inequality, which replaces \eqref{eq:cross}:
\begin{equation}\label{eq:wcross}
\frac{p_{i\tau}}{p_{J\tau}}\ \ge\ \kappa_J\ \ge\ \sigma_\Lambda(u).
\end{equation}

Since every $\hat a_\tau$ is nonnegative, it suffices to find, on every sample path, one arrival $\tau$ where $\hat a_\tau\ge V$. There are three cases.

\emph{(a) $\alg$ matches $i$ at an arrival $\tau$ that is not a release arrival.} Then $\hat a_\tau=a_\tau=1\ge V$, because $v_t(u)\le1$ for every $t$.

\emph{(b) $\alg$ matches $i$ at a release arrival $\tau$.} Then $z_\tau\le p_{i\tau}<p_{J\tau}$, so $\hat a_\tau=p_{i\tau}/p_{J\tau}\ge\sigma_\Lambda(u)$ by \eqref{eq:wcross}. Let $t\in T_i$. If $t\le\tau$, then $i$ is available at $t$, so $v_t(u)=\sigma_\Lambda(u)$. If $t>\tau$, then, because $\tau$ is a release arrival and $(i,\tau)$ succeeds on this path, condition~(ii) of Definition~\ref{def:wrelarr} says that $\alg$ probes at $t$ a resource whose reduced price is at least $r_ip_{it}(1-g(u))$, so again $v_t(u)=\sigma_\Lambda(u)$. Hence $V\le\sigma_\Lambda(u)\le\hat a_\tau$.

\emph{(c) $\alg$ never matches $i$.} Then $i$ is available at every arrival, so $v_t(u)=\sigma_\Lambda(u)$ for every $t$. If $\alg$ never releases, then $T_i=\emptyset$ and $V=0$ by Corollary~\ref{cor:norelease}. Otherwise let $\tau$ be the last arrival at which $\alg$ releases. We claim that $\tau$ is a release arrival. Condition~(i) holds at $\tau$, since $p_{i\tau}<z_\tau\le p_{J\tau}$. For condition~(ii), compare after $\tau$ the run on $\bz$ (run~1) with the run on $\bz^{(\tau)}$ (run~2). The two runs are identical before $\tau$, both probe $i$ at $\tau$, which fails in run~1 and succeeds in run~2, and they see the same thresholds after $\tau$, from the available sets $S\cup\{i\}$ and $S$ with $i\notin S$. Run~1 never matches $i$ and does not release after $\tau$, so Lemma~\ref{lem:shadow} applies. Take $t>\tau$ with $(i,t)\in E$ and $z_t\le p_{it}$. Run~1 does not probe $i$ at $t$, since the probe would succeed. As $i$ is available, run~1 probes a resource $k\ne i$ whose reduced price is at least $r_ip_{it}(1-g(u))$, and by Lemma~\ref{lem:shadow} run~2 probes the same $k$. This is condition~(ii). So $\tau$ is a release arrival with $z_\tau\le p_{J\tau}$, and $\hat a_\tau=p_{i\tau}/p_{J\tau}\ge\sigma_\Lambda(u)\ge V$ by \eqref{eq:wcross}.
\end{proof}

\section*{Acknowledgements}
We thank Ali Aouad and Santiago Balseiro for encouraging us to continue pursuing the general case of OMSR.   

\section*{AI Disclosure}
Claude Opus 5 and 5.5 (Anthropic) were used to explore ideas, proof strategies, and help with writing. We started the interaction with typed notes that included (i) the algorithm, main technical ideas, and a proof sketch for the unweighted setting and (ii) several high-level ideas for extending results to the weighted setting. Formal results were derived in the subsequent interaction. In addition, the parameters of the perturbation functions in Appendix~\ref{app:numerics} (Table~\ref{tab:bands}) were optimized numerically by the AI. We have verified all results in this manuscript and take full responsibility for its content and correctness.

\bibliographystyle{alphanat}
\bibliography{STOC_v5}

\end{document}